%% file: main.tex
\documentclass[11pt]{article}

\usepackage[T1]{fontenc}
\usepackage[margin=1in]{geometry}
\usepackage{microtype}
\usepackage{graphicx}
\usepackage{amsmath,amssymb}
\usepackage[numbers,sort&compress]{natbib}
\usepackage[hidelinks]{hyperref}

\title{Score-based stochastic reduced-order models of barotropic quasi-geostrophic turbulence}
\author{Ludovico Theo Giorgini\\[0.4em]
\normalsize Department of Mathematics, Massachusetts Institute of Technology\\
\normalsize Cambridge, MA 02139, USA\\[0.3em]
\normalsize\href{mailto:ludogio@mit.edu}{ludogio@mit.edu}}
\date{}

\begin{document}
\maketitle

\begin{abstract}
In many turbulent flows, the quantities of interest are coarse-grained variables, either because resolving the fine scales is computationally prohibitive or because only coarse observations are available; in both cases, data-driven reduced-order models that evolve only these variables are required. Even when the governing equations are known, their projection onto the coarse variables is not closed, and the central difficulty lies in representing the feedback of the unresolved scales so that the reduced model reproduces the statistical and dynamical properties of the projected high-resolution system. We address this problem with the score-based framework of Refs.~\cite{Giorgini2026Score,Giorgini2026Conditional}, constructing a reduced-order model of quasi-geostrophic turbulence whose stationary distribution coincides by construction with the distribution learned from data and whose dynamics are calibrated to reproduce the finite-time correlations of the energies in individual wavenumber shells. The resulting model reproduces the stationary statistics of the resolved flow with high accuracy, and its calibration is computationally inexpensive because it requires no forward integration of the model.
\end{abstract}

\section{Introduction}

Turbulent flows in the oceans, atmosphere, and engineering systems couple nonlinear processes across disparate spatial and temporal scales. The quantities of scientific interest, however, often depend only on a much smaller set of coarse observables and their statistics rather than on the complete high-dimensional state. The goal is therefore to use coarse data from observations, experiments, or high-fidelity simulations to construct efficient stochastic models that evolve only these resolved variables while reproducing the statistical and dynamical observables of interest. The difficulty is that the dynamics of the coarse variables are generally not closed: unresolved degrees of freedom influence their evolution through effective dissipation, fluctuations, feedback, and memory~\cite{Berner2017,WoutersLucarini2013}. Closure modeling addresses this problem by constructing a self-contained reduced model that accounts for unresolved effects, using available data and physical structure.

Many closure methods proceed from the bottom up: they infer or prescribe a detailed local evolution law from instantaneous tendencies, short-lag transitions, or other local relationships, and then integrate that law to determine whether it reproduces the stationary statistics and finite-time observables of interest. This creates a structural mismatch: a model flexible enough to represent the local dynamics may contain many degrees of freedom, whereas the targets constrain only particular combinations of them; long-time fidelity is obtained only if fitting the local behavior recovers the combinations that matter for those targets. Different approaches balance long-time fidelity against model and calibration cost. Markovian \emph{a priori} closures are calibrated from local data and use only the current coarse state~\cite{Frezat2022}; they are efficient, but can miss unresolved feedback and memory, while local errors can accumulate under integration~\cite{WoutersLucarini2013,Frezat2022}. Non-Markovian models add temporal histories or colored noise~\cite{WoutersLucarini2013}, whereas \emph{a posteriori} methods simulate the reduced model during calibration and adjust its parameters to reduce discrepancies between the resulting observables and their data-derived targets. These strategies can improve fidelity, but require either greater model complexity or expensive optimization, often involving repeated forward simulations and differentiation through the model~\cite{Frezat2022}. Achieving long-time statistical and dynamical fidelity from a bottom-up description of local dynamics therefore remains difficult and expensive.

Recent approaches pursue the opposite direction: using observed statistical and dynamical properties to constrain the construction of dynamical models, including learning dynamics from invariant measures~\cite{BotvinickGreenhouse2023} and constructing score-based stochastic models from stationary statistics and finite-time dynamical information~\cite{Giorgini2026Score,Giorgini2026Conditional}. Related work uses response theory to assess reduced models and to infer additional constraints on their interactions~\cite{FalascaZanna2026}. Here, we consider the score-based modeling approach developed in Refs.~\cite{Giorgini2026Score,Giorgini2026Conditional} and use it to construct statistical closures for turbulence. The main idea of this method is to begin with the target observables and determine which stochastic models satisfy the corresponding statistical and dynamical constraints. This leaves the local evolution law initially unspecified and typically yields an infinite family of models with different local dynamics compatible with the imposed constraints. This admissible family is determined directly from the observable constraints, without repeated simulations of candidate closures.

The central mathematical object enabling this construction is the score function, defined as the gradient of the logarithm of a probability density. Recent advances in generative score modeling make it possible to estimate this quantity directly from high-dimensional samples, without reconstructing and normalizing the underlying density~\cite{Song2021}. The stationary score characterizes a broad class of stochastic models that preserve the observed steady-state density by construction; combining it with linear response theory through the generalized fluctuation--dissipation theorem~\cite{Marconi2008,Giorgini2024Response,Giorgini2025Response} provides complementary dynamical constraints that restrict this class using selected finite-time dynamical observables~\cite{Giorgini2026Score,Giorgini2026Conditional}. Together, these constraints yield a practical construction of Markovian stochastic closures with prescribed stationary statistics and selected finite-time observables as calibration targets.

In this work, we apply this method to construct and test statistical closures for quasi-geostrophic (QG) turbulence, a simplified description of turbulent motion in rapidly rotating fluids that is central to understanding large-scale atmospheric and oceanic circulation~\cite{MarstonTobias2023}. Building statistical closures for this system is challenging because unresolved eddies influence the formation and persistence of large-scale flows: a reduced model must capture both their long-time statistics and their temporal variability~\cite{FrederiksenKitsiosOKane2024}. Its geophysical relevance and demanding closure problem make QG turbulence a useful testbed for the proposed method.

\section{The quasi-geostrophic turbulence system}
\label{sec:qg-system}

Quasi-geostrophic (QG) dynamics describe the slow evolution of rotating flows whose leading-order horizontal force balance is between pressure gradients and the Coriolis force. This balance is central to large-scale atmospheric and oceanic motion. QG models retain the interactions responsible for the evolution of vortices, the propagation of Rossby waves, and the organisation of turbulent motion into large-scale currents. For a review, see Ref.~\cite{MarstonTobias2023}.

We consider the barotropic QG model, in which the horizontal velocity is independent of depth. The flow evolves on a doubly periodic horizontal domain, with coordinates $x$ and $y$ representing the eastward and northward directions, respectively. Planetary rotation is represented through the beta-plane approximation,
\begin{equation}
 f(y)=f_0+\beta y,
\end{equation}
where $f$ is the Coriolis parameter, $f_0$ is its value at a reference latitude, and $\beta$ is its northward gradient. This approximation retains the variation of planetary rotation with latitude within a planar description of the flow.

The horizontal velocity $\boldsymbol{u}=(u,v)$ is expressed through a streamfunction $\psi(x,y,t)$ as
\begin{equation}
 u=-\partial_y\psi,
 \qquad
 v=\partial_x\psi.
\end{equation}
This representation satisfies incompressibility, $\nabla\cdot\boldsymbol{u}=0$. The relative vorticity,
\begin{equation}
 \zeta=\partial_x v-\partial_y u=\Delta\psi,
 \qquad
 \Delta=\partial_x^2+\partial_y^2,
\end{equation}
measures the local rotation of the fluid relative to the rotating reference frame. The streamfunction, and hence the velocity, can be recovered from the vorticity by solving this Poisson equation with periodic boundary conditions.

The forced and dissipative barotropic QG equation is
\begin{equation}
 \partial_t\zeta
 +J(\psi,\zeta)
 +\beta\,\partial_x\psi
 =-\mu\zeta+\nu\Delta\zeta+\xi(x,y,t),
 \label{eq:qg}
\end{equation}
where
\begin{equation}
 J(\psi,\zeta)
 =\partial_x\psi\,\partial_y\zeta
 -\partial_y\psi\,\partial_x\zeta
 =\boldsymbol{u}\cdot\nabla\zeta.
\end{equation}

The time derivative $\partial_t\zeta$ describes the local evolution of relative vorticity. The Jacobian $J(\psi,\zeta)$ represents its advection by the flow. Because the advecting velocity itself depends on $\zeta$, this term is nonlinear and couples different spatial scales.

The term $\beta\,\partial_x\psi=\beta v$ represents the advection of planetary vorticity by northward or southward motion. Together, the two advective terms express transport of $q=\zeta+\beta y$, the barotropic potential vorticity up to an additive constant. In the absence of forcing and dissipation, $q$ is conserved along fluid trajectories. When a fluid parcel moves northward or southward, it encounters a different planetary vorticity and must change its relative vorticity to conserve $q$. The resulting velocity changes couple the motion of neighboring parcels, allowing disturbances to propagate as Rossby waves. Because planetary vorticity varies in the north--south direction but not in the east--west direction, the dynamics depend on the direction of motion.

On the right-hand side, $-\mu\zeta$ represents linear drag, with drag coefficient $\mu\geq0$. It damps the velocity field and provides a sink for energy transferred towards large scales. The viscous term $\nu\Delta\zeta$, with kinematic viscosity $\nu\geq0$, smooths vorticity gradients and preferentially damps small spatial scales.

The forcing $\xi(x,y,t)$ is a zero-mean Gaussian field, white in time and correlated in space. Writing $\boldsymbol r=(x,y)$, we define it through the Wiener increments
\begin{equation}
 \xi(\boldsymbol r,t)\,\mathrm dt=\mathrm dW_f(\boldsymbol r,t)
 =A_f\sum_{\boldsymbol k\in\mathcal K_f}
 \sqrt{S_f(\boldsymbol k)}\,e^{i\boldsymbol k\cdot\boldsymbol r}\,\mathrm dW_{\boldsymbol k}(t),
 \label{eq:forcing-definition}
\end{equation}
where $\mathcal K_f$ is the symmetric set of nonzero forced wavevectors. The complex Wiener processes are independent for one member of each conjugate pair, satisfy $\mathbb E|\mathrm dW_{\boldsymbol k}|^2=\mathrm dt$, and obey $\mathrm dW_{-\boldsymbol k}=\mathrm dW_{\boldsymbol k}^*$ so that the forcing is real. Its spectrum and amplitude are
\begin{equation}
 S_f(\boldsymbol k)=\exp\!\left[-\frac{(|\boldsymbol k|-k_f)^2}{2(\Delta k_f)^2}\right],
 \qquad
 A_f=\left[\frac{2\varepsilon_f}
 {\displaystyle\sum_{\boldsymbol k\in\mathcal K_f}S_f(\boldsymbol k)/|\boldsymbol k|^2}\right]^{1/2}.
 \label{eq:forcing-amplitude}
\end{equation}
Here $k_f$ sets the characteristic forcing scale and $\Delta k_f$ its spectral bandwidth. Both sums include both members of each conjugate pair. The amplitude $A_f$ is chosen so that the forcing injects kinetic energy at mean rate $\varepsilon_f$. For a fixed spectral shape, $A_f$ therefore scales as $\sqrt{\varepsilon_f}$.

The resulting dynamics combine nonlinear transfer between scales, wave propagation, energy injection, and dissipation. Their balance can sustain a statistically stationary turbulent regime in which instantaneous flow fields continue to fluctuate while their probability distribution remains unchanged. This provides the setting for constructing a closure that reproduces the stationary statistics and temporal evolution of the resolved flow.

\section{Closure modelling for quasi-geostrophic turbulence}
\label{sec:conventional-closure}

Constructing a closure for QG turbulence requires representing the exchange of physical quantities across the boundary between resolved and unresolved scales. In the inviscid, unforced barotropic system, nonlinear advection conserves both kinetic energy and enstrophy. These conservation laws constrain the transfer between scales: energy tends to move towards larger scales, while enstrophy tends to move towards smaller scales. Removing small-scale degrees of freedom therefore changes both the dissipation of resolved fluctuations and the transfer of energy back into the resolved flow. The planetary-vorticity gradient adds a further difficulty by supporting Rossby waves and promoting anisotropic interactions between eddies and zonal currents. A closure must represent these competing effects without explicitly evolving the unresolved motion.

We define the resolved variables through a spectral cutoff $K$. Let $P_K$ denote the projection onto Fourier modes with spatial wavenumber magnitude no greater than $K$, and write
\begin{equation}
 \overline{\zeta}=P_K\zeta,
 \qquad
 \overline{\psi}=P_K\psi,
 \qquad
 \overline{\zeta}=\Delta\overline{\psi}.
\end{equation}
The cutoff determines the smallest spatial scales retained by the reduced model. Applying this projection to the full QG equation gives
\begin{equation}
 \partial_t\overline{\zeta}
 +P_KJ(\overline{\psi},\overline{\zeta})
 +\beta\,\partial_x\overline{\psi}
 =-\mu\overline{\zeta}
 +\nu\Delta\overline{\zeta}
 +P_K\xi+\mathcal{R}_K,
 \label{eq:projected-qg}
\end{equation}
where
\begin{equation}
 \mathcal{R}_K
 =P_KJ(\overline{\psi},\overline{\zeta})
 -P_KJ(\psi,\zeta)
\end{equation}
is the unresolved contribution to the resolved vorticity tendency. This term depends on interactions involving the discarded modes and cannot, in general, be determined from the instantaneous resolved state alone. Setting $\mathcal{R}_K=0$ neglects the contribution of unresolved interactions to the evolution of the resolved flow.

As a conventional closure, we consider a spectral eddy-viscosity model supplemented by stochastic backscatter. The combination of dissipative and stochastic contributions is an established approach to geophysical subgrid modelling~\cite{FrederiksenDavies1997}, reviewed recently by Frederiksen, Kitsios and O'Kane~\cite{FrederiksenKitsiosOKane2024}. In this model, the unresolved tendency is approximated by
\begin{equation}
 \mathcal{R}_K\approx-\mathcal{D}_K\overline{\zeta}+\xi_{\mathrm{bs}},
\end{equation}
where $\mathcal{D}_K$ is a positive spectral damping operator and $\xi_{\mathrm{bs}}$ is an additional stochastic forcing representing backscatter.

The eddy-viscosity term represents the drain of resolved fluctuations into unresolved scales. Its action on a retained Fourier coefficient is
\begin{equation}
 \widehat{\mathcal{D}_K\overline{\zeta}}_{\boldsymbol{k}}
 =\nu_{\mathrm e}\,|\boldsymbol{k}|^2
 \left(\frac{|\boldsymbol{k}|}{K}\right)^p
 \widehat{\overline{\zeta}}_{\boldsymbol{k}},
 \label{eq:eddy-viscosity}
\end{equation}
where $\nu_{\mathrm e}\geq0$ controls the damping amplitude and $p>0$ determines its concentration near the cutoff. The hat denotes a Fourier coefficient and $\boldsymbol{k}$ its wavevector. This scale dependence preferentially removes fluctuations near the smallest retained scales, limiting their artificial accumulation while reducing direct damping of the largest structures.

The stochastic backscatter term represents energy returned from unresolved motion through a zero-mean Gaussian forcing, white in time and supported on the retained modes. With $\mathcal K_K=\{\boldsymbol k\in\mathbb Z^2:0<|\boldsymbol k|\leq K\}$, we use
\begin{equation}
 \xi_{\mathrm{bs}}(\boldsymbol r,t)\,\mathrm dt
 =A_{\mathrm{bs}}\sum_{\boldsymbol k\in\mathcal K_K}
 \sqrt{S_{\mathrm{bs}}(\boldsymbol k)}\,e^{i\boldsymbol k\cdot\boldsymbol r}
 \,\mathrm dW^{\mathrm{bs}}_{\boldsymbol k}(t),
 \label{eq:backscatter-definition}
\end{equation}
with
\begin{equation}
 S_{\mathrm{bs}}(\boldsymbol k)
 =\exp\!\left[-\frac{(|\boldsymbol k|-k_{\mathrm{bs}})^2}{2(\Delta k_{\mathrm{bs}})^2}\right],
 \qquad
 A_{\mathrm{bs}}=\left[\frac{2\varepsilon_{\mathrm{bs}}}
 {\displaystyle\sum_{\boldsymbol k\in\mathcal K_K}S_{\mathrm{bs}}(\boldsymbol k)/|\boldsymbol k|^2}\right]^{1/2}.
 \label{eq:backscatter-amplitude}
\end{equation}
Here $k_{\mathrm{bs}}$ and $\Delta k_{\mathrm{bs}}$ specify the centre and bandwidth of the backscatter spectrum, and $\varepsilon_{\mathrm{bs}}$ is its mean kinetic-energy injection rate. The Wiener processes $W^{\mathrm{bs}}_{\boldsymbol k}$ are independent of the physical forcing and independent across representatives of distinct conjugate pairs, with $\mathbb E|\mathrm dW^{\mathrm{bs}}_{\boldsymbol k}|^2=\mathrm dt$ and $\mathrm dW^{\mathrm{bs}}_{-\boldsymbol k}=(\mathrm dW^{\mathrm{bs}}_{\boldsymbol k})^*$. Both sums include both members of each conjugate pair. This additional forcing models fluctuations generated by the eliminated degrees of freedom.

For prescribed damping and backscatter spectral shapes, their amplitudes are calibrated from the energy and enstrophy budgets of the projected high-resolution data. The resolved quantities are
\begin{equation}
 E_K=\frac12\left\langle|\nabla\overline{\psi}|^2\right\rangle,
 \qquad
 Z_K=\frac12\left\langle\overline{\zeta}^{\,2}\right\rangle,
\end{equation}
where $\langle\cdot\rangle$ denotes a spatial average. The reference trajectories determine the mean contributions that the closure must supply to these two budgets after accounting for resolved drag, viscosity, and physical forcing. The damping removes energy and enstrophy in proportions determined by its spectral shape and the observed spectrum; the stochastic forcing injects them in proportions determined by its covariance. For each prescribed choice of $p$, $k_{\mathrm{bs}}$, and $\Delta k_{\mathrm{bs}}$, matching the two mean budgets gives two algebraic equations for the damping amplitude $\nu_{\mathrm e}$ and the backscatter energy injection rate $\varepsilon_{\mathrm{bs}}$. We prescribe a narrow, finite backscatter bandwidth to represent energy returned over neighbouring resolved scales. The chosen width spreads the injection across several wavenumber shells while keeping it concentrated around $k_{\mathrm{bs}}$. Fixing this width leaves the characteristic scale of energy return, set by $k_{\mathrm{bs}}$, to be selected from the reference data. We select $p$ and $k_{\mathrm{bs}}$ by testing candidate pairs: for each pair, fit the amplitudes from the two budgets, simulate the resulting reduced model, and compare its stationary and temporal statistics with those of the reference data.

The resulting reduced model is Markovian: its drift depends only on the current resolved field, and its stochastic increments are independent of the past. The calibration has only two free parameters, $p$ and $k_{\mathrm{bs}}$; the two amplitudes follow from the budget equations. This small search space keeps calibration inexpensive even though candidate models must be integrated. We use this model as a baseline because eddy viscosity with stochastic backscatter is widely used in geophysical turbulence modelling~\cite{FrederiksenKitsiosOKane2024}, has the same Markovian structure as our score-based model, and is straightforward to calibrate.

\section{Score-based reduced stochastic modelling}
\label{sec:score-model}

The score-based approach constructs a reduced stochastic model directly from the stationary statistics of the resolved variables. Let $\boldsymbol{x}\in\mathbb{R}^n$ denote the resolved state, consisting here of the independent real and imaginary components of the retained Fourier coefficients. We denote its target stationary probability density by $p_{\mathrm{ss}}(\boldsymbol{x})$ and define the stationary score as
\begin{equation}
 \boldsymbol{s}(\boldsymbol{x})
 =\nabla_{\boldsymbol{x}}\log p_{\mathrm{ss}}(\boldsymbol{x}).
 \label{eq:stationary-score}
\end{equation}
All derivatives in this section are taken with respect to the resolved state. The score describes how the probability density varies locally in state space and can be learned directly from stationary samples.

Following Refs.~\cite{Giorgini2026Score,Giorgini2026Conditional}, we consider the It\^o stochastic differential equation
\begin{equation}
\begin{aligned}
 \mathrm{d}\boldsymbol{x}_t
 ={}&\left[
 \bigl(D(\boldsymbol{x}_t)+R(\boldsymbol{x}_t)\bigr)\boldsymbol{s}(\boldsymbol{x}_t)
 +\nabla\cdot\bigl(D(\boldsymbol{x}_t)+R(\boldsymbol{x}_t)\bigr)
 \right]\mathrm{d}t\\
 &+\sqrt{2}\,\Sigma(\boldsymbol{x}_t)\,\mathrm{d}\boldsymbol{W}_t,
\end{aligned}
\label{eq:score-rom}
\end{equation}
where $\boldsymbol{W}_t$ is a standard $n$-dimensional Wiener process. The matrices satisfy
\begin{equation}
 D=D^{\mathsf T}\succeq0,
 \qquad R=-R^{\mathsf T},
 \qquad \Sigma\Sigma^{\mathsf T}=D.
\end{equation}
Thus $D$ is symmetric and positive semidefinite, while $R$ is antisymmetric. Their sum $M=D+R$ is the mobility matrix. The divergence term accounts for its state dependence and vanishes when $D$ and $R$ are constant.

The stationary density is preserved by construction. Substituting $p_{\mathrm{ss}}$ into the Fokker--Planck equation, the identity $p_{\mathrm{ss}}\boldsymbol{s}=\nabla p_{\mathrm{ss}}$ makes the drift contribution associated with $D$ cancel the diffusive contribution. The remaining stationary probability current is
\begin{equation}
 \boldsymbol{J}_{\mathrm{ss}}=\nabla\cdot(Rp_{\mathrm{ss}}),
 \qquad \nabla\cdot\boldsymbol{J}_{\mathrm{ss}}=0.
 \label{eq:stationary-current}
\end{equation}
Its divergence vanishes because $R$ is antisymmetric and mixed derivatives commute. Hence $p_{\mathrm{ss}}$ satisfies the stationary Fokker--Planck equation for every admissible choice of $D$ and $R$ in this construction.

The matrices $D$ and $R$ control different aspects of the motion while preserving the same stationary density. The matrix $D$ sets the covariance of the stochastic increments,
\begin{equation}
 \operatorname{Cov}\!\left[
 \mathrm{d}\boldsymbol{x}_t\mid\boldsymbol{x}_t=\boldsymbol{x}
 \right]=2D(\boldsymbol{x})\,\mathrm{d}t,
\end{equation}
and supplies the associated drift $D\boldsymbol{s}+\nabla\cdot D$. Together, these terms describe the reversible, dissipative component of the dynamics. They determine the directions and strengths of random fluctuations and contribute to their relaxation times.

The matrix $R$ supplies a deterministic contribution $R\boldsymbol{s}+\nabla\cdot R$ that can sustain a nonzero stationary probability current. It represents circulation through state space without changing the stationary density, allowing the model to describe nonreversible dynamics. When $R$ is constant, $R\boldsymbol{s}$ is perpendicular to the score and therefore tangent to surfaces of constant probability density. When $R$ depends on the state, the additional drift $\nabla\cdot R$ ensures that the resulting probability current remains divergence-free and preserves $p_{\mathrm{ss}}$.

The stationary score therefore specifies a family of reduced models with a common invariant density. Once the score has been learned and fixed, the remaining reduced-order modelling problem is to estimate $D$ and $R$ from dynamical information in the data. Their calibration determines how the model moves through the prescribed distribution, including the persistence, oscillation, and decorrelation of resolved fluctuations. We will use the calibration methods developed in the two papers~\cite{Giorgini2026Score,Giorgini2026Conditional}, incorporating information about the physical QG system to constrain the form and estimation of $D$ and $R$, without repeated candidate-model integrations.

In the following, we describe how the stationary score and the matrices $D$ and $R$ are constructed from the reference data. We first estimate the score, which specifies the stationary distribution preserved by the reduced model. We then use temporal statistics and information about the physical QG system to determine the dissipative and circulatory components of its dynamics.

\subsection{Estimation of the stationary score}
\label{sec:score-estimation}

We estimate the stationary score from snapshots of high-resolution QG trajectories projected onto the resolved Fourier modes. The independent Fourier components are centred and rescaled using statistics of the reference data, so that differences in their amplitudes do not dominate the learning problem. Denoting the resulting coordinates by $\boldsymbol{z}$, we construct a model of their stationary density and obtain its score by differentiation.

The density model combines a Gaussian reference with a learned nonlinear correction:
\begin{equation}
 p_\theta(\boldsymbol{z})
 =\frac{1}{Z_\theta}\,
 q_{\mathrm G}(\boldsymbol{z})
 \exp\!\bigl[h_\theta(\boldsymbol{z})\bigr],
 \label{eq:learned-density}
\end{equation}
where $q_{\mathrm G}$ is a Gaussian density whose covariance is estimated from the reference data, $h_\theta$ is a scalar function represented by a neural network, and $Z_\theta$ is the normalising constant. The Gaussian reference captures the covariance structure of the resolved fluctuations. The neural correction represents departures from Gaussian statistics, including dependencies between modes that cannot be described by covariance alone.

We constrain the correction to be bounded by writing
\begin{equation}
 h_\theta(\boldsymbol{z})
 =B\tanh\!\left(\frac{f_\theta(\boldsymbol{z})-f_\theta(\boldsymbol{z}_{\mathrm{ref}})}{B}\right),
 \label{eq:bounded-correction}
\end{equation}
where $f_\theta$ is the scalar network output, $B>0$ controls the maximum magnitude of the correction, and $\boldsymbol{z}_{\mathrm{ref}}$ is the mean of the Gaussian reference. Subtracting the network output at this fixed state sets $h_\theta(\boldsymbol{z}_{\mathrm{ref}})=0$ and makes the correction independent of an arbitrary additive constant in $f_\theta$. Since $|\tanh u|\leq1$, we have $|h_\theta|\leq B$. Consequently, the Gaussian reference controls the decay of the density at large amplitudes, and the learned correction cannot destroy its normalisability.

The hyperbolic tangent also preserves the flexibility of the correction when its argument is small: $\tanh u=u+O(u^3)$. Thus, wherever $|f_\theta(\boldsymbol{z})-f_\theta(\boldsymbol{z}_{\mathrm{ref}})|\ll B$,
\begin{equation}
 h_\theta(\boldsymbol{z})\approx f_\theta(\boldsymbol{z})-f_\theta(\boldsymbol{z}_{\mathrm{ref}}),
 \qquad
 \nabla h_\theta(\boldsymbol{z})\approx\nabla f_\theta(\boldsymbol{z}).
\end{equation}
In this regime, the network learns the log-density correction directly, with little modification by the bounding function. Near minima of the effective potential $-\log p_\theta$, which correspond to maxima of the density, this allows the local shape to be learned directly whenever the correction remains in the linear regime. For larger arguments, the hyperbolic tangent smoothly limits the correction while retaining a differentiable density model. The score is
\begin{equation}
 \boldsymbol{s}_\theta(\boldsymbol{z})
 =\nabla_{\boldsymbol{z}}\log q_{\mathrm G}(\boldsymbol{z})
 +\nabla_{\boldsymbol{z}}h_\theta(\boldsymbol{z}).
 \label{eq:learned-score}
\end{equation}
The unknown normalising constant disappears under differentiation. Learning a scalar correction also ensures that the resulting score is the gradient of a well-defined log-density, so that the reduced stochastic model can preserve the learned probability density as described above.

The neural network evaluates the correction using spatial fields reconstructed from the retained Fourier coefficients. A convolutional architecture with periodic boundary treatment processes these fields at several spatial scales and combines the resulting features into a scalar output. This representation allows the model to learn spatial organisation and interactions across scales while respecting the periodic geometry of the QG system.

Training uses denoising score matching. A reference state $\boldsymbol{z}$ is perturbed by independent Gaussian noise,
\begin{equation}
 \widetilde{\boldsymbol{z}}
 =\boldsymbol{z}+\sigma\boldsymbol{\epsilon},
 \qquad
 \boldsymbol{\epsilon}\sim\mathcal{N}(0,I),
\end{equation}
and the network parameters are fitted by minimising
\begin{equation}
 \mathcal{L}(\theta)
 =\mathbb{E}\!\left[
 \left\|
 \sigma\boldsymbol{s}_\theta
 (\boldsymbol{z}+\sigma\boldsymbol{\epsilon})
 +\boldsymbol{\epsilon}
 \right\|^2
 \right].
 \label{eq:denoising-score-loss}
\end{equation}
To evaluate this loss during training, we draw snapshots from the reference trajectories, add independently generated Gaussian perturbations to them, and average the squared error over the resulting examples. The perturbations provide a known training target, allowing the score to be estimated without evaluating the data density or its derivatives. At the population optimum, this objective recovers the score of the reference distribution convolved with a Gaussian of width $\sigma$. The method therefore learns a smooth approximation to the stationary density; the smoothing scale controls the balance between regularity and preservation of fine statistical structure.

\subsection{Construction and calibration of the diffusion and circulation matrices}
\label{sec:mobility-calibration}

The learned stationary density determines the distribution of resolved flow states, but it does not determine how long fluctuations persist or how they evolve between spatial scales. Our goal is to construct $D$ and $R$ to reproduce both the observed short-time dynamics and the longer-time correlations of a selected set of physically relevant observables. A natural choice for these observables is the kinetic energy contained at different spatial scales, together with the total resolved energy. Energy transfer between scales is central to QG turbulence: it governs how turbulent motion supplies and sustains large-scale flows, while forcing and dissipation control the overall energy balance. We therefore group Fourier modes of similar wavenumber magnitude into shells and use their energies to measure how the intensity of the resolved motion is distributed across scales. The autocorrelation of each shell energy measures how long fluctuations at those scales persist, while cross-correlations between shell energies describe how fluctuations at different scales are related over time. The total resolved energy complements this description by measuring variations in the overall intensity of the flow. Matching these energy correlations constrains the time scales of the resolved dynamics: two flows can have the same mean energy at each scale but differ in how rapidly their energy fluctuations decay. Energy correlations do not, however, describe all aspects of the motion. A Fourier mode can change its phase without changing its energy, so energy measurements alone cannot determine this part of its evolution. We therefore also use the real and imaginary parts of selected Fourier coefficients when fitting the circulation matrix $R$.

The calibration follows the approach developed in Refs.~\cite{Giorgini2026Score,Giorgini2026Conditional}: the score fixes the stationary density, while observed temporal statistics constrain the mobility. We first estimate a constant mobility from short-time data, then introduce state-dependent matrices to reproduce the finite-time correlations of the selected observables.

We begin with the constant-mobility ansatz
\begin{equation}
 \Phi=D_0+R_0,
 \qquad
 \mathrm{d}\boldsymbol{x}
 =\Phi\boldsymbol{s}(\boldsymbol{x})\,\mathrm{d}t
 +\sqrt{2D_0}\,\mathrm{d}\boldsymbol{W}.
\end{equation}
Let $\boldsymbol{y}=\boldsymbol{x}-\mathbb{E}_{\mathrm{HF}}[\boldsymbol{x}]$ denote centred resolved coordinates, where $\mathbb{E}_{\mathrm{HF}}$ denotes an average over the reference QG data. Define
\begin{equation}
\begin{aligned}
 G(t)&=\mathbb{E}_{\mathrm{HF}}[\boldsymbol{y}_t\boldsymbol{y}_0^{\mathsf T}],\\
 S&=\mathbb{E}_{\mathrm{HF}}[\boldsymbol{s}(\boldsymbol{x}_0)\boldsymbol{y}_0^{\mathsf T}].
\end{aligned}
\end{equation}
For the constant model, the initial covariance derivative satisfies $\dot G(0^+)=\Phi S$. We therefore estimate
\begin{equation}
 \widehat\Phi=\widehat{\dot G}(0^+)\,\widehat S^\dagger,
 \label{eq:constant-mobility-estimate}
\end{equation}
where the covariance derivative is obtained from a short-lag fit and ${}^\dagger$ denotes a pseudoinverse. Retaining the measured matrix $\widehat S$ accounts for the approximate learned score; for an exact score evaluated under its own density, $S=-I$. Further details of this short-time construction are given in the two papers.

We average $\widehat\Phi$ over the translation and reflection symmetries of the QG system. In physical Fourier coordinates, the resulting matrix is block diagonal, with one block for each independent complex Fourier mode:
\begin{equation}
 \widehat\Phi_k=
 \begin{pmatrix}d_k^\Phi&-r_k^\Phi\\r_k^\Phi&d_k^\Phi\end{pmatrix}.
\end{equation}
This structure reflects the spatial homogeneity of the system: translating a flow pattern within the periodic domain does not change its statistical properties. The real and imaginary parts of a Fourier mode represent the same spatial pattern shifted by a quarter wavelength, so neither component should be preferred. The equal diagonal entries $d_k^\Phi$ assign them the same diffusion strength, while the antisymmetric entries $r_k^\Phi$ allow circulation in their two-dimensional plane, representing a directional contribution to the mode's phase evolution. A fixed coupling between distinct wavevectors would generally depend on their phases relative to the chosen spatial origin and is therefore absent from the symmetry-averaged constant matrix. This does not imply that the Fourier modes evolve independently: their interactions can still enter through the joint score. The reflection symmetry further assigns equal coefficients to nonzonal modes with wavevectors $(k_x,k_y)$ and $(k_x,-k_y)$. For zonal modes, a nonzero rotation coefficient would select a preferred direction of meridional phase motion, which this reflection symmetry excludes.

The symmetric and antisymmetric parts of the symmetry-averaged mobility estimate $\widehat\Phi$ give preliminary diffusion and circulation estimates, respectively. Their sampling uncertainties are estimated by repeatedly drawing collections of complete reference trajectories with replacement and recalculating the coefficients. These estimates constrain the short-time coordinate dynamics, but do not constrain the model to reproduce the finite-time correlations of the target observables.

To reproduce the observed finite-time energy correlations, we allow the diffusion matrix $D$ and the circulation matrix $R$ to depend on the resolved state, using the constant estimates $D_0$ and $R_0$ as a reference for reproducing the observed short-time dynamics. We distinguish motion that changes total resolved energy from motion that redistributes it among the resolved modes. The first represents changes in the overall intensity of the flow, while the second represents internal rearrangement. We assume that circulation contributes to this internal rearrangement: nonlinear interactions among retained modes exchange energy between them, while Rossby-wave propagation changes their phases, and neither process changes their total energy. Diffusion is allowed to contain both types of motion. State-dependent coefficients let us enforce this distinction at each flow configuration. The subsequent fit uses the preliminary estimate from fitting the constant mobility to the observed short-time dynamics as a reference, with a penalty on changes in the mean diffusion and additional short-time constraints on both the diffusion matrix $D$ and the circulation matrix $R$, as described below.

Write the resolved kinetic energy as
\begin{equation}
 E(\boldsymbol{x})=\frac12\boldsymbol{x}^{\mathsf T}H\boldsymbol{x},
 \qquad \boldsymbol{g}(\boldsymbol{x})=\nabla E=H\boldsymbol{x},
\end{equation}
where the diagonal matrix $H$ contains the Fourier weights defining kinetic energy. Let $C$ and $\Lambda$ be positive diagonal matrices, with equal entries for the real and imaginary components of each mode. The matrix $C$ controls background diffusion, while $\Lambda$ controls internal fluctuations.

To constrain those internal fluctuations, define
\begin{equation}
\begin{aligned}
 \boldsymbol{u}&=\Lambda\boldsymbol{g},
 &q&=\boldsymbol{g}^{\mathsf T}\Lambda\boldsymbol{g},\\
 \Pi(\boldsymbol{x})&=\frac{qI-\boldsymbol{u}\boldsymbol{g}^{\mathsf T}}{q+\varepsilon},
\end{aligned}
\end{equation}
where $\varepsilon>0$ prevents division by zero near vanishing energy. Since $\boldsymbol{g}^{\mathsf T}\Pi=0$, the vector $\Pi\boldsymbol{v}$ is tangent to a surface of constant energy for any vector $\boldsymbol{v}$.

The diffusion and circulation matrices are then
\begin{equation}
\begin{aligned}
 D(\boldsymbol{x})&=C+\Pi(\boldsymbol{x})\Lambda\Pi(\boldsymbol{x})^{\mathsf T},\\
 R(\boldsymbol{x})&=\Pi(\boldsymbol{x})A_{\boldsymbol r}\Pi(\boldsymbol{x})^{\mathsf T},
\end{aligned}
\label{eq:energy-mobility}
\end{equation}
where $A_{\boldsymbol r}$ consists of antisymmetric Fourier-pair blocks
\begin{equation}
 (A_{\boldsymbol r})_k=\begin{pmatrix}0&-r_k\\r_k&0\end{pmatrix}.
\end{equation}

The unknown parameters are the diagonal coefficients of $C$ and $\Lambda$, and the rotation coefficients $r_k$, with coefficients required to be equal by the spatial symmetries fitted as a single parameter.

This construction makes $D$ positive definite and $R$ antisymmetric. Moreover,
\begin{equation}
 (D-C)\nabla E=0,\qquad R\nabla E=0.
\end{equation}
Including the divergence terms in the stochastic drift, both internal components preserve total energy.

We fit $C$ and $\Lambda$ using energy correlations. Let
\begin{equation}
 \boldsymbol{e}(\boldsymbol{x})=
 \bigl(E_1(\boldsymbol{x}),\ldots,E_m(\boldsymbol{x}),E(\boldsymbol{x})\bigr)^{\mathsf T}
\end{equation}
contain the shell energies and their sum. Thus its component $e_i$ equals the shell energy $E_i$ for $1\leq i\leq m$, and $e_{m+1}=E$ is total energy. From projected high-resolution trajectories, we estimate
\begin{equation}
 K^{\mathrm{HF}}_{ij}(t)=\mathbb{E}_{\mathrm{HF}}\left[
 \bigl(e_i(\boldsymbol{x}_t)-\overline e_i^{\mathrm{HF}}\bigr)
 \bigl(e_j(\boldsymbol{x}_0)-\overline e_j^{\mathrm{HF}}\bigr)\right],
\end{equation}
where $\overline e_i^{\mathrm{HF}}$ is the reference mean. In practice, the average uses pairs of saved states separated by the lag $t$. The normalised reference correlations are
\begin{equation}
 \rho^{\mathrm{HF}}_{ij}(t)=
 \frac{K^{\mathrm{HF}}_{ij}(t)}{\sqrt{K^{\mathrm{HF}}_{ii}(0)K^{\mathrm{HF}}_{jj}(0)}}.
\end{equation}
The diffusion fit uses their symmetric part,
\begin{equation}
 \rho^{\mathrm{tar}}(t)=\frac{\rho^{\mathrm{HF}}(t)+\rho^{\mathrm{HF}}(t)^{\mathsf T}}{2}.
\end{equation}

The rationale for calibrating $D$ is to constrain the score-based model to reproduce the linear responses of the reference system for the selected observables and perturbations. The generalized fluctuation--dissipation theorem (GFDT) expresses these responses in terms of statistics of the unperturbed stationary system. It therefore allows response information to be estimated from available data without performing additional perturbed simulations. The identity developed in Ref.~\cite{Giorgini2026Conditional} connects these response constraints to derivatives of lagged correlations and to the unknown mobility. Here, we evaluate this connection through a finite-dimensional approximation of the conditional evolution of observables. This yields an efficient calibration of $D$ based on stationary averages and matrix calculations, without repeated integrations of candidate reduced models. Ref.~\cite{Giorgini2026Conditional} provides the response-theoretic derivation and further details. 

To approximate the evolution of energy observables, we augment the vector of individual modal energies with nonlinear functions of the observed state. After subtracting the stationary mean of each component under $p_\theta$, we denote the augmented vector by $\Psi=(\Psi_1,\ldots,\Psi_N)^{\mathsf T}$. The purpose of this augmentation is to obtain a collection of observables whose conditional expectations can be approximated by a closed linear evolution. This approximation is used for calibration; the reduced stochastic model evolves the retained Fourier coefficients.

For a candidate stationary score-based model, the conditional expectations evolve under the linear semigroup $\mathcal K_t$. We define this operator and the covariance matrices of the augmented observables by
\begin{equation}
\begin{aligned}
 (\mathcal K_tF)(\boldsymbol{x})
 &=\mathbb{E}[F(\boldsymbol{x}_t)\mid\boldsymbol{x}_0=\boldsymbol{x}],\\
 K_{\Psi,ab}(t)
 &=\mathbb{E}_{p_\theta}[\Psi_a(\boldsymbol{x}_t)\Psi_b(\boldsymbol{x}_0)],
 \qquad B=K_\Psi(0).
\end{aligned}
\end{equation}
The matrix $K_\Psi(t)$ contains their lagged covariances, and $B$ is their stationary covariance matrix. Below, we approximate the action of $\mathcal K_t$ on $\Psi$ by a finite-dimensional matrix acting on the same augmented vector.

Writing $M=D+R$, the GFDT identity gives
\begin{equation}
 \frac{\mathrm d}{\mathrm dt}K_{\Psi,ab}(t)
 =-\mathbb{E}_{p_\theta}\!\left[
 \nabla(\mathcal K_t\Psi_a)^{\mathsf T}M\nabla\Psi_b
 \right].
 \label{eq:energy-gfdt-identity}
\end{equation}
All gradients and coefficient matrices on the right-hand side are evaluated at the initial state. This identity expresses the correlation derivative as the linear response of $\Psi_a$ to an impulsive drift perturbation in the direction $-M\nabla\Psi_b$~\cite{Giorgini2026Conditional}. At zero lag, $\mathcal K_0\Psi_a=\Psi_a$. For the chosen Fourier-pair circulation and observables that depend only on modal energies,
\begin{equation}
 \mathbb{E}_{p_\theta}[\nabla\Psi_a^{\mathsf T}R\nabla\Psi_b]=0.
\end{equation}
Consequently, the initial covariance derivative is determined by diffusion:
\begin{equation}
 Q_{ab}(D)
 :=\mathbb{E}_{p_\theta}[\nabla\Psi_a^{\mathsf T}D\nabla\Psi_b]
 =-\dot K_{\Psi,ab}(0^+).
 \label{eq:energy-diffusion-form}
\end{equation}
Here, $\dot K_{\Psi,ab}(0^+)$ is the initial rate of change of the covariance predicted by the score-based model with diffusion matrix $D$. We use $Q(D)$ below to approximate the model's finite-time energy correlations and calibrate $D$ by comparing these predictions with correlations measured in the reference simulations.

The same matrix can be expressed using the diffusion contribution to the backward generator. For a smooth observable $F$, define
\begin{equation}
 \mathcal L_DF
 =\frac{1}{p_\theta}\nabla\cdot(p_\theta D\nabla F).
\end{equation}
Integration by parts, with vanishing boundary terms, gives
\begin{equation}
 -\mathbb{E}_{p_\theta}[\Psi_a\mathcal L_D\Psi_b]
 =\mathbb{E}_{p_\theta}[\nabla\Psi_a^{\mathsf T}D\nabla\Psi_b]
 =Q_{ab}(D).
\end{equation}
Thus $Q(D)$ represents the action of diffusion on the observable collection. Its evaluation requires the gradients of the observables and the diffusion matrix at stationary samples. The matrix $B$ is evaluated from the same samples and remains fixed during calibration.

To obtain finite-time predictions, we approximate the conditional evolution within the span of the chosen functions:
\begin{equation}
 (\mathcal K_t\Psi)(\boldsymbol{x})\approx A(t)\Psi(\boldsymbol{x}),
 \qquad A(0)=I.
 \label{eq:energy-semigroup-approximation}
\end{equation}
Here $A(t)$ is a matrix of time-dependent coefficients. This approximation gives $K_\Psi(t)\approx A(t)B$. Substitution into Eq.~\eqref{eq:energy-gfdt-identity}, using the vanishing circulation contribution within this collection, yields the closed matrix equation
\begin{equation}
 \dot A(t)B=-A(t)Q(D).
\end{equation}
On a linearly independent set of centred functions, for which $B$ is invertible, its solution gives
\begin{equation}
 K_\Psi^{\mathrm{pred}}(t)
 =\exp[-Q(D)B^{-1}t]B.
 \label{eq:energy-correlation-prediction}
\end{equation}
This is a Galerkin approximation of the stochastic Koopman generator~\cite{Klus2020}. Its finite-time accuracy depends on how well the chosen functions represent the evolving conditional expectations. Although $R$ makes no contribution to the projected matrix, it can influence the full model's energy correlations through functions outside this finite collection.

In Ref.~\cite{Giorgini2026Conditional}, the gradient $\nabla(\mathcal K_t\Psi_a)$ is evaluated through the conditional score of the transition density, estimated from observed lagged pairs. Here, Eq.~\eqref{eq:energy-semigroup-approximation} supplies a finite-dimensional approximation of this evolution. The resulting calibration therefore uses the same GFDT identity without estimating a conditional score. For each candidate $D$, stationary samples determine $Q(D)$, and Eq.~\eqref{eq:energy-correlation-prediction} gives the predicted correlations without integrating new stochastic trajectories.

Because shell and total energies are linear combinations of modal energies, we define a fixed matrix $O$ that performs these combinations. Each shell row sums the modal energies belonging to that shell, and the total-energy row sums all retained modal energies. The columns corresponding to additional nonlinear functions have zero entries; any rescaling of the modal functions is included in the corresponding coefficients of $O$. Thus
\begin{equation}
 \boldsymbol{e}-\mathbb{E}_{p_\theta}[\boldsymbol{e}]=O\Psi.
\end{equation}
Consequently,
\begin{equation}
 K^{\mathrm{pred}}(t)=O\exp[-Q(D)B^{-1}t]\,B\,O^{\mathsf T},
\end{equation}
and
\begin{equation}
 \rho^{\mathrm{pred}}_{ij}(t)=
 \frac{K^{\mathrm{pred}}_{ij}(t)}{\sqrt{K^{\mathrm{pred}}_{ii}(0)K^{\mathrm{pred}}_{jj}(0)}}.
\end{equation}
These predictions depend on the diffusion coefficients and permit fitting $D$ before the circulation coefficients are determined.

We determine $C$ and $\Lambda$ by minimising
\begin{equation}
\begin{aligned}
 \mathcal J_D={}&\operatorname{mean}_{i,\ell}
 \left[\rho^{\mathrm{pred}}_{ii}(t_\ell)-\rho^{\mathrm{tar}}_{ii}(t_\ell)\right]^2\\
 &+\operatorname{mean}_{i,j,\ell}
 \left[\rho^{\mathrm{pred}}_{ij}(t_\ell)-\rho^{\mathrm{tar}}_{ij}(t_\ell)\right]^2\\
 &+\frac{\alpha}{2}\operatorname{mean}_a
 \left[\frac{\overline d_a-d_a^\Phi}{u_a^\Phi}\right]^2.
\end{aligned}
\label{eq:diffusion-objective}
\end{equation}
Here $t_\ell$ are the comparison lags, $a$ labels groups of Fourier coefficients required to share the same parameters by symmetry, $\overline d_a$ is the corresponding mean diagonal entry of $D$, and $d_a^\Phi$ and $u_a^\Phi$ are its preliminary estimate and uncertainty from the constant-mobility calculation. The first two terms measure autocorrelation and full correlation-matrix errors. The last term discourages departures from the short-time estimate that are not justified by the finite-time data. The coefficient $\alpha\geq0$ sets the relative weight of this penalty in the fit.

We minimise $\mathcal J_D$ subject to a constraint on the short-time fluctuations of total energy. Let $D_f$ denote half the covariance per unit time of the retained Fourier coefficients of the stochastic forcing $\xi$ introduced in Section~\ref{sec:qg-system}.

The stochastic forcing produces short-time fluctuations in the total resolved energy. Half their mean-square growth rate is computed directly from the known forcing covariance and stationary reference snapshots:
\begin{equation}
 q_f=\mathbb{E}_{\mathrm{HF}}[\nabla E^{\mathsf T}D_f\nabla E].
\end{equation}
For the reduced model, the corresponding quantity is
\begin{equation}
 q_E(C)=\mathbb{E}_{p_\theta}[\nabla E^{\mathsf T}C\nabla E].
\end{equation}
This expectation is evaluated by averaging over stationary samples from $p_\theta$ for each candidate $C$, without integrating the reduced model. Only $C$ contributes because the internal terms preserve energy. We require $q_E(C)$ to lie within an interval $[q_f^-,q_f^+]$ representing uncertainty in the reference value.

The diffusion coefficients are therefore obtained from the single constrained problem
\begin{equation}
\begin{aligned}
 &\min_{C,\Lambda}\mathcal J_D(C,\Lambda)\\
 &\text{subject to}\quad C>0,\qquad\Lambda>0,\qquad q_f^-\leq q_E(C)\leq q_f^+.
\end{aligned}
\end{equation}

After fixing $C$ and $\Lambda$, we fit the circulation coefficients. The observables $F_b$ include the energies and the real and imaginary parts of selected nonzonal Fourier coefficients. Energy depends on the squared modulus of a Fourier coefficient and is unchanged when its phase rotates. We therefore also consider observables that resolve the phase. Writing $a_k=\operatorname{Re}\widehat\zeta_k$ and $b_k=\operatorname{Im}\widehat\zeta_k$, both centred, this information is contained in
\begin{equation}
 \Delta_k(t)=\mathbb{E}_{\mathrm{HF}}[b_k(t)a_k(0)]
 -\mathbb{E}_{\mathrm{HF}}[a_k(t)b_k(0)].
\end{equation}
We select modes for which this directional signal is distinguishable from sampling uncertainty and whose correlations persist over time and agree between separate sets of reference trajectories. Modes connected by meridional reflection are included together. Their real and imaginary components provide the phase information needed to fit circulation.

From the reference trajectories, we calculate
\begin{equation}
 Y_{b,h}(t)=\frac{F_b(\boldsymbol{x}_{t+h})-F_b(\boldsymbol{x}_{t-h})}{2h}.
\end{equation}
For a stationary Markov diffusion,
\begin{equation}
 \lim_{h\downarrow0}\mathbb{E}[Y_{b,h}(t)\mid\boldsymbol{x}_t=\boldsymbol{x}]
 =\nabla F_b^{\mathsf T}\bigl(R\boldsymbol{s}+\nabla\cdot R\bigr).
\end{equation}
Thus centred temporal differences identify the contribution of the stationary probability current. Estimates at several short lags are extrapolated to zero lag to reduce finite-lag bias, producing targets $\widehat Y_b$.

With $\Lambda$ fixed, we express the circulation matrix as a linear combination of known matrices. Each independent coefficient $r_a$ is assigned to a non-zonal mode and its meridionally reflected counterpart, with both modes sharing the same value. Zonal rotation coefficients vanish by symmetry. We denote the corresponding set of modes by $I_a$; it contains a single mode when $k_y=0$, since reflection then leaves the wavevector unchanged. The corresponding matrix $J^{(a)}$ is block diagonal, with a block for each real--imaginary Fourier pair:
\begin{equation}
 \bigl(J^{(a)}\bigr)_k=
 \begin{cases}
 \begin{pmatrix}0&-1\\1&0\end{pmatrix}, & k\in I_a,\\[4pt]
 0_{2\times2}, & k\notin I_a.
 \end{cases}
\end{equation}
Thus, on each mode in $I_a$, $J^{(a)}$ maps the pair of real and imaginary components $(x_{k,\mathrm{Re}},x_{k,\mathrm{Im}})$ to $(-x_{k,\mathrm{Im}},x_{k,\mathrm{Re}})$, the direction of an infinitesimal phase rotation. It gives zero on all other modes. Applying the energy-preserving projection $\Pi$ gives the state-dependent matrix
\begin{equation}
 R^{(a)}(\boldsymbol{x})=\Pi(\boldsymbol{x})J^{(a)}\Pi(\boldsymbol{x})^{\mathsf T}.
\end{equation}
Then
\begin{equation}
 R(\boldsymbol{x})=\sum_a r_aR^{(a)}(\boldsymbol{x}).
\end{equation}
The functions $R^{(a)}$ are known after fitting $\Lambda$; the remaining unknowns are the scalar coefficients $r_a$.

Define
\begin{equation}
 A_{ba}(\boldsymbol{x})=\nabla F_b(\boldsymbol{x})^{\mathsf T}
 \left[R^{(a)}(\boldsymbol{x})\boldsymbol{s}(\boldsymbol{x})
 +\nabla\cdot R^{(a)}(\boldsymbol{x})\right].
\end{equation}
The predicted current is then $\sum_a A_{ba}r_a$, linear in the unknown coefficients. We determine them by minimising
\begin{equation}
\begin{aligned}
 \mathcal J_R(\boldsymbol r)
 ={}&\mathbb{E}_{\mathrm{HF}}\left[\sum_b w_b\left(
 \sum_a A_{ba}(\boldsymbol{x}_t)r_a-\widehat Y_b(t)\right)^2\right]\\
 &+\gamma\sum_a\left(\frac{r_a-r_a^\Phi}{u_a^R}\right)^2.
\end{aligned}
\label{eq:circulation-objective}
\end{equation}
The weights $w_b$ balance energy and phase observables using their measured temporal-derivative variances. Here $r_a^\Phi$ denotes the preliminary rotation coefficient and $u_a^R$ its estimated uncertainty. The penalty discourages departures from the preliminary circulation estimate that are not supported by the short-time data. The coefficient $\gamma\geq0$ sets the relative weight of this penalty in the fit.

The calibration combines short-time dynamical constraints with energy correlations to determine $D$ and estimates of the stationary probability current to determine $R$. Throughout this procedure, the learned stationary density remains fixed: it is encoded by the score and preserved by the model for every admissible choice of $D$ and $R$. These matrices can therefore be calibrated to reproduce dynamical observables without altering the prescribed stationary statistics. Both fits are performed offline, using reference data, samples from the learned density, and the finite-dimensional correlation approximation. No integration of candidate reduced models is required during calibration, avoiding the computational cost of repeated stochastic simulations.

\section{Results}
\label{sec:results}

We integrated the full QG equations and projected the resulting trajectories onto the Fourier modes with $|\boldsymbol k|\leq K=16$. We refer to these projected reference trajectories as the high-fidelity (HF) data. We calibrated and integrated the QG model with eddy viscosity and stochastic backscatter, denoted the low-fidelity (LF) model, and the proposed score-based model (SM). Both reduced models evolve the same resolved Fourier modes. The numerical integration, parameter values, and calibration settings are specified in Appendix~\ref{app:implementation}.

\input{results-figures}

\section{Conclusions}
\label{sec:conclusions}

We have constructed a stochastic reduced model for the large-scale component of forced, dissipative barotropic quasi-geostrophic turbulence on a beta plane, defined by the Fourier modes of the flow below a prescribed spectral cutoff. The construction consists of two steps. First, the probability density of the large-scale flow states is learned from snapshots of high-resolution simulations by training a neural network to estimate its score, that is, the gradient of the logarithm of the density. Second, the learned score defines the drift of a stochastic differential equation whose stationary distribution coincides, by construction, with the learned density. This property holds for every admissible choice of the remaining coefficients of the equation, so that the stationary statistics of the model are fixed independently of the calibration of its dynamics. These coefficients form two matrices that determine how the model evolves in time within the learned distribution. The diffusion matrix sets the intensity of the random forcing acting on each variable and the correlations between the random forcings of different variables, and thereby controls how rapidly fluctuations are generated and damped. The circulation matrix adds a deterministic drift that circulates through the space of flow states without altering the learned distribution, and describes nonreversible motion such as the propagation of Rossby waves and the phase evolution of Fourier modes. Both matrices are calibrated against temporal correlations measured in the reference simulations, subject to physical constraints: the symmetries of the flow, the requirement that part of the dynamics redistribute energy among scales without changing the total energy, as the nonlinear interactions do, and the magnitude of the energy fluctuations injected by the forcing. The calibration requires only averages over the reference data and over samples from the learned density, without repeated integration of candidate models.

By construction, the stationary statistics of the model are those of the learned density, and numerical integrations of the model confirm this property: its stationary statistics are in nearly perfect agreement with those of the high-resolution simulations projected onto the same large scales. These include the mean energy spectrum and its variability; the probability distributions of total energy, enstrophy, zonal energy fraction, and vorticity kurtosis; the non-Gaussian statistics of velocity and vorticity increments; and the joint distributions of these quantities. The conventional reduced model with eddy viscosity and stochastic backscatter, calibrated on the same data, exhibits substantial biases in several of these statistics. This result addresses a central difficulty of turbulence closure: a coarse-grained model must retain the statistical organisation of the resolved flow after the unresolved degrees of freedom have been eliminated. In reduced models that prescribe a local evolution law, such as the eddy-viscosity model considered here, the stationary statistics emerge only after that law has been integrated, and errors in the local dynamics can produce statistical biases. In the present approach, the learned joint distribution of the retained variables is instead imposed as a structural constraint on the model, so that its stationary statistics are determined primarily by the accuracy of the learned score and of the numerical time integration, independently of the calibration of the dynamics.

The dynamics are calibrated separately, within the family of models that share the learned density. The diffusion parameters are fitted, using the generalized fluctuation--dissipation identities developed in Ref.~\cite{Giorgini2026Conditional}, to reproduce the finite-time correlations of the shell energies and the total energy, with individual modal energies and nonlinear functions forming the observable library used in the Galerkin approximation. This calibration of the dynamics is performed without integrating candidate models. The model reproduces the structure of the time correlations of the shell energies, although in some cases less accurately, with the correlations of some of the largest-scale shells decaying too slowly. For observables not used in the calibration, the accuracy is mixed: the correlations of enstrophy, vorticity kurtosis, and zonal shear decay too slowly, those of the zonal energy fraction and of the increment flatness decay too rapidly, and the oscillatory correlation of the Fourier mode with wavevector $(1,1)$ and the negative short-lag correlation of the eddy-to-jet energy transfer are not reproduced. Both these correlations and those of the calibrated observables could be improved by augmenting the library of observables used in the fit. Determining how the number of free parameters in the diffusion and circulation matrices limits which and how many correlations can be reproduced simultaneously is left to future work.

A natural next step is the extension of this reduced model to two-layer baroclinic quasi-geostrophic turbulence on a beta plane, a standard idealised model of oceanic and atmospheric flows that retains the vertical structure responsible for baroclinic instability and eddy heat transport~\cite{GalletFerrari2021,QiMajda2016}. This extension would broaden the physical scope of the reduced model to baroclinic oceanic and atmospheric dynamics. The potential implications for Earth-system modelling are considerable. Many processes that control the variability of the ocean and atmosphere, such as mesoscale ocean eddies, remain unresolved in climate models and must be represented through parameterizations whose errors affect the simulated climate statistics~\cite{Berner2017}. Reduced models whose stationary statistics are constrained by construction to match high-resolution data, and whose dynamics are calibrated against temporal correlations without repeated simulations of candidate models, could serve as stochastic emulators that retain the non-Gaussian statistics and variability of the resolved flow at a substantially reduced computational cost. Their use as stochastic parameterizations would additionally require specifying their coupling to the resolved equations of a host model. Such models would enable large ensembles for uncertainty quantification, long integrations for the study of low-frequency variability, and systematic exploration of the response of the climate system to changes in forcing. The results presented here for barotropic turbulence constitute a first step in this direction.

\section*{Acknowledgments}
The author thanks Keaton Burns for useful discussions.

\input{appendix}

\end{document}

%% file: results-figures.tex
\begin{figure}[p]
\centering
\includegraphics[width=\textwidth]{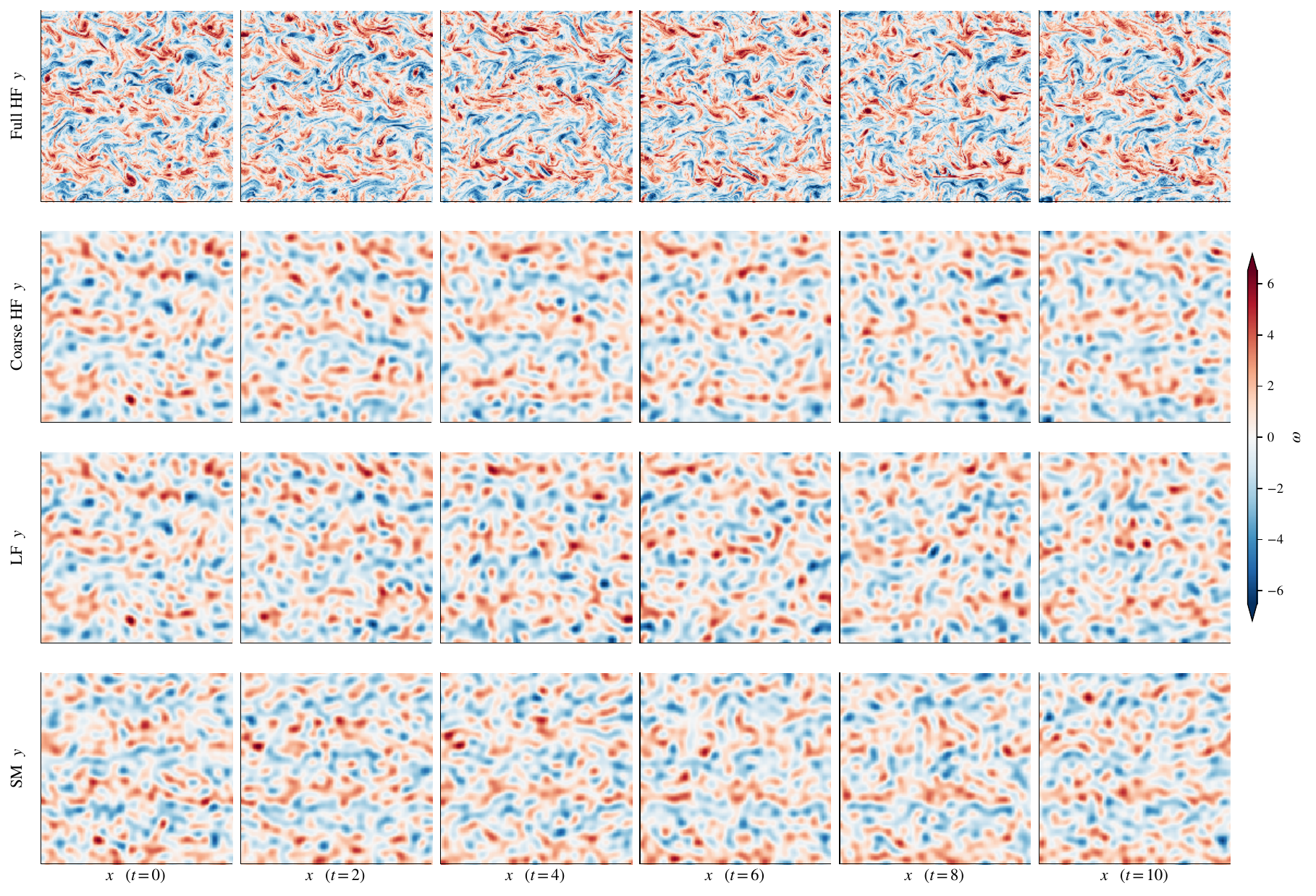}
\caption{Vorticity snapshots. Rows, from top to bottom: full HF, coarse HF, LF, and SM. Columns, from left to right: $t=0,2,4,6,8,10$. Full and coarse HF show the same realization; SM shows an independent realization. All panels share the colour scale.}
\label{fig:flow-sequences}
\end{figure}
\begin{figure}[p]
\centering
\includegraphics[width=\textwidth]{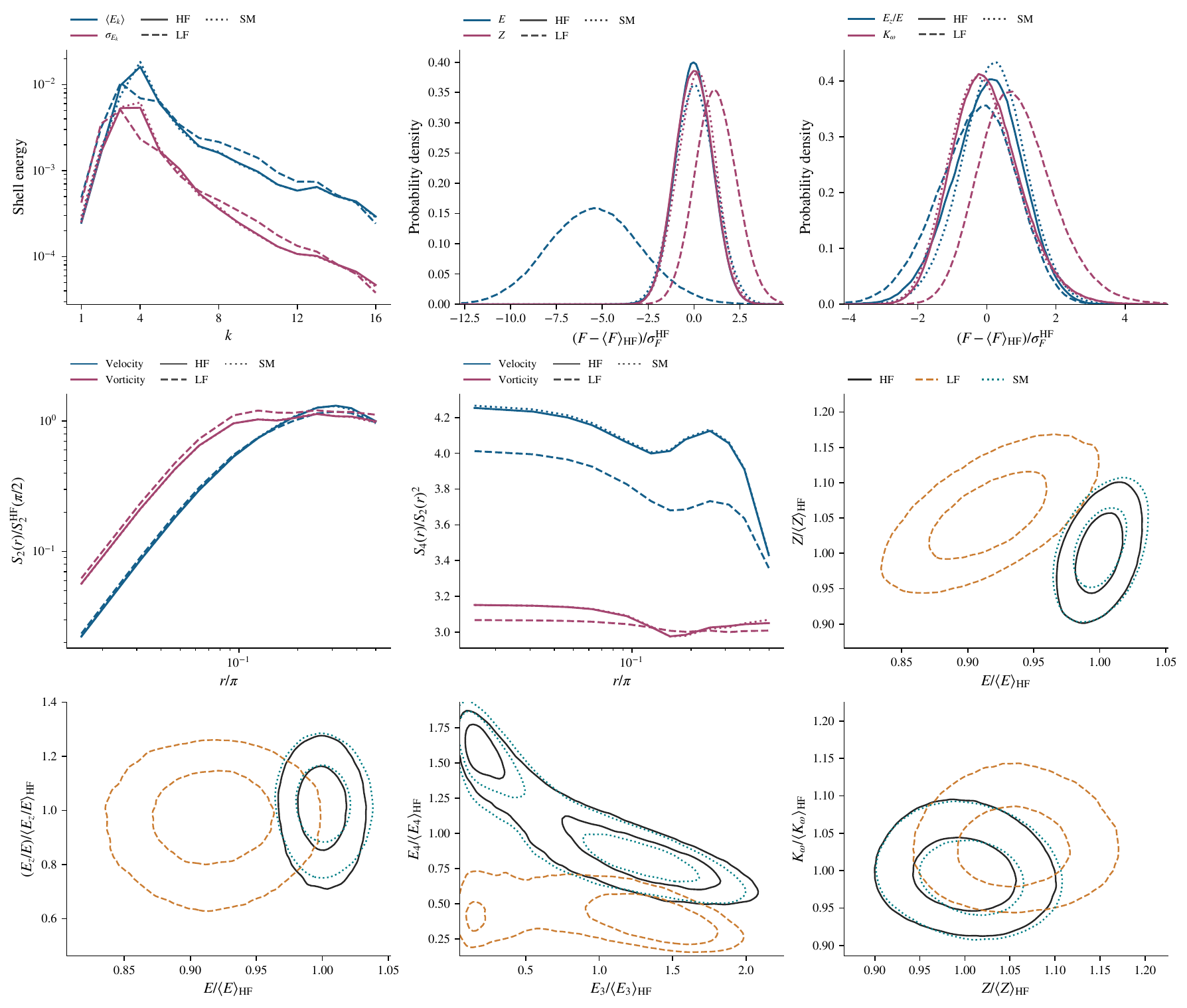}
\caption{Statistical comparisons of HF, LF, and SM. In each row, panels are listed from left to right. Top: shell-energy means and standard deviations; PDFs of total energy and enstrophy; PDFs of zonal energy fraction and vorticity kurtosis. Middle: second-order velocity and vorticity structure functions; increment flatness $S_4/S_2^2$; joint energy--enstrophy PDF. Bottom: joint PDFs of energy and zonal fraction, $E_3$ and $E_4$, and enstrophy and vorticity kurtosis. Contours enclose 50\% and 90\% of the displayed probability.}
\label{fig:statistics}
\end{figure}
\begin{figure}[p]
\centering
\includegraphics[width=\textwidth]{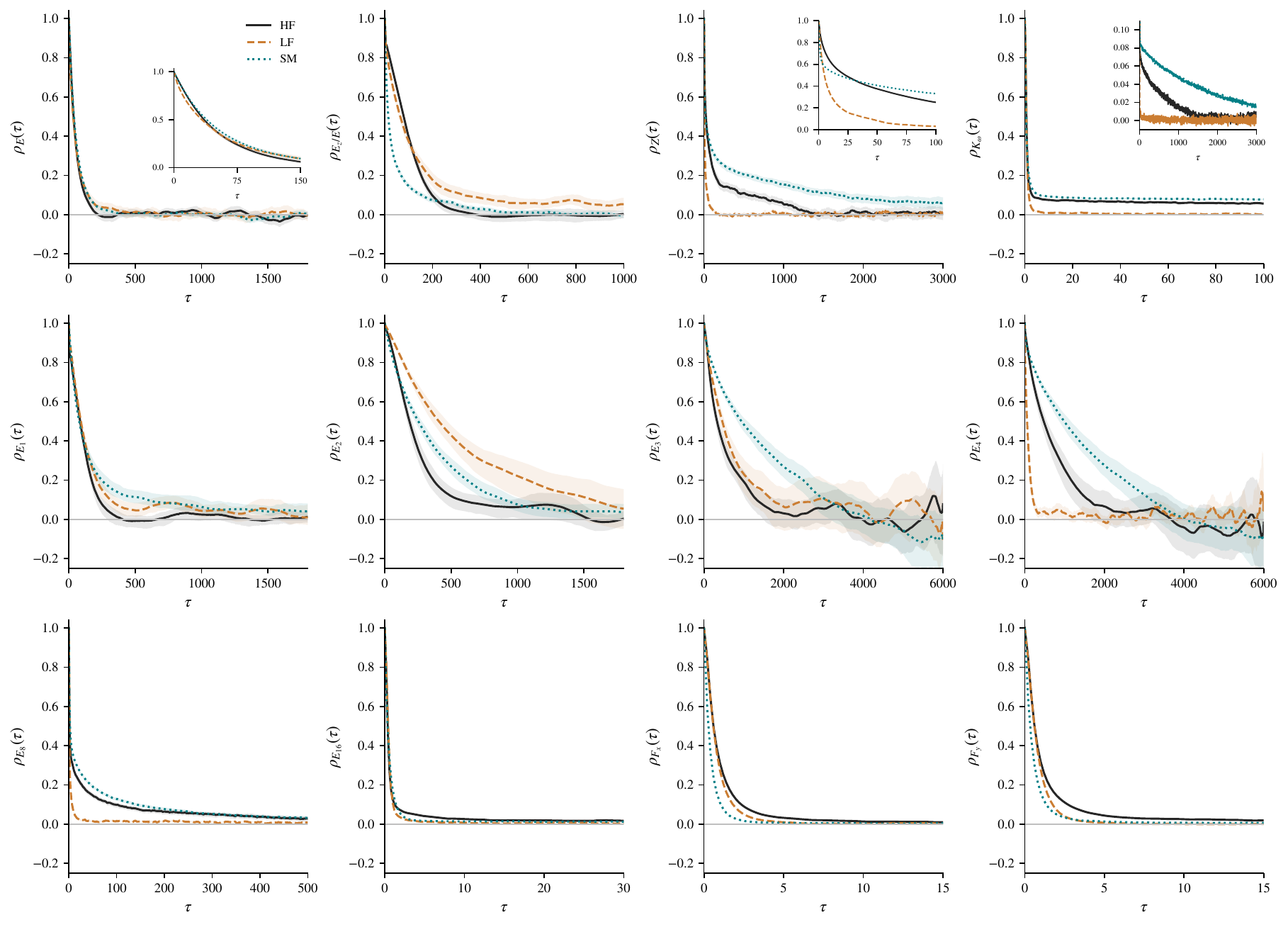}
\caption{Normalized autocorrelations for HF, LF, and SM. Top row, left to right: total energy, zonal energy fraction, enstrophy, and vorticity kurtosis. Middle row: $E_1,E_2,E_3,E_4$. Bottom row: $E_8,E_{16}$ and vorticity-increment flatness in the $x$ and $y$ directions at separation $\pi/16$. Insets show short-time energy and enstrophy correlations and the long-time kurtosis correlation. Shading indicates 95\% bootstrap confidence intervals.}
\label{fig:acf-comparison}
\end{figure}
\begin{figure}[p]
\centering
\includegraphics[width=\textwidth]{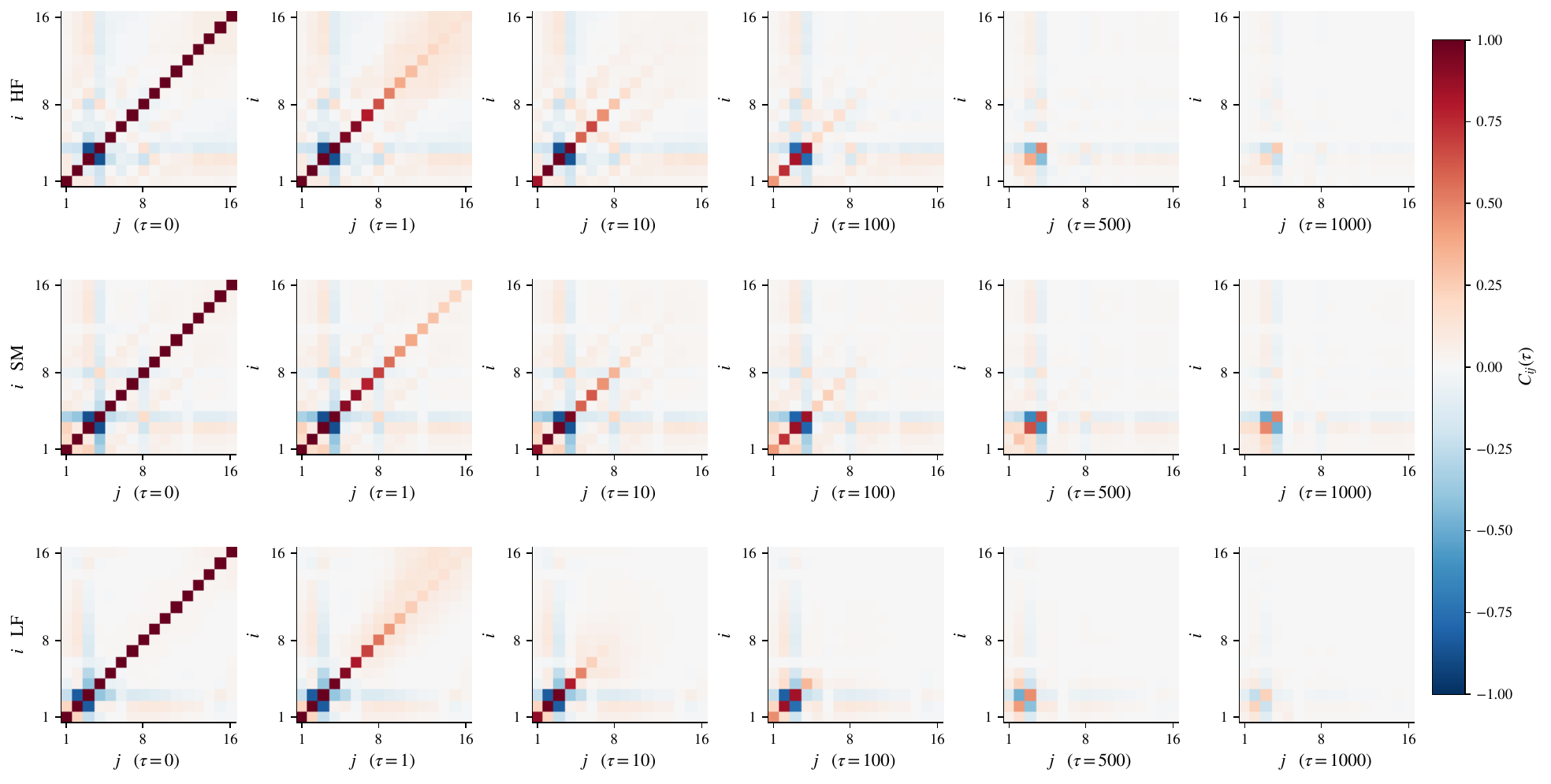}
\caption{Normalized shell-energy cross-correlation matrices. Rows, from top to bottom: HF, SM, and LF. Columns, from left to right: $\tau=0,1,10,100,500,1000$. Entry $(i,j)$ correlates $E_i(t+\tau)$ with $E_j(t)$. All panels share the colour scale.}
\label{fig:cross-correlations}
\end{figure}
\begin{figure}[p]
\centering
\includegraphics[width=\textwidth]{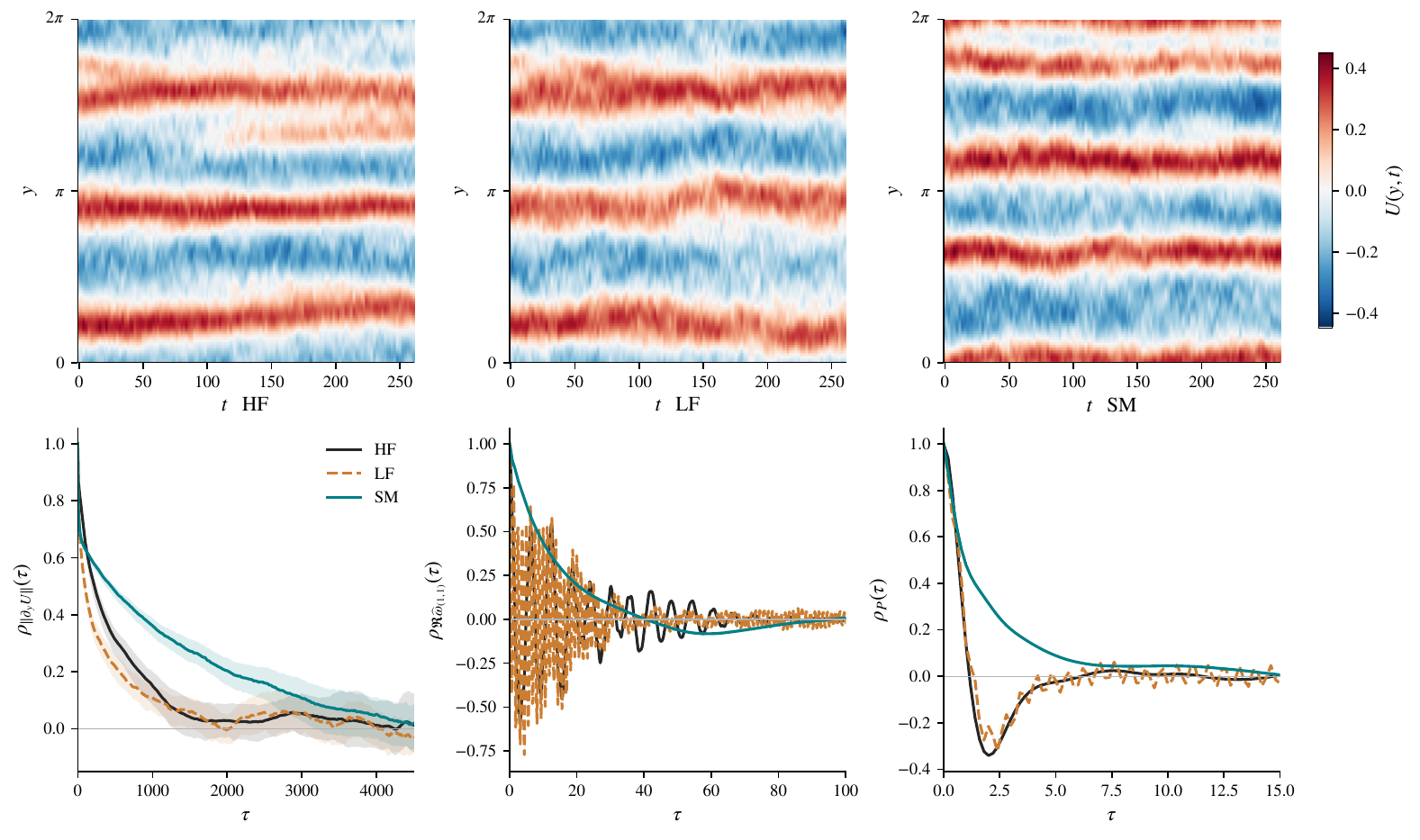}
\caption{Zonal-flow evolution and temporal correlations. Top row, left to right: zonal velocity $U(y,t)$ for HF, LF, and SM. Bottom row, left to right: normalized autocorrelations of root-mean-square zonal shear, $\Re\widehat\omega_{(1,1)}$, and the QG eddy-to-jet transfer observable $P$. Shading indicates 95\% bootstrap confidence intervals.}
\label{fig:jet-dynamics}
\end{figure}

Figures~\ref{fig:flow-sequences}--\ref{fig:jet-dynamics} compare the spatial evolution, one-time statistics, and temporal correlations of the three models. Unless explicitly labelled ``Full HF'', the reference fields and observables are evaluated after projection to the resolved modes. The one-time SM statistics in Fig.~\ref{fig:statistics} are estimated from an ensemble of 240 trajectories over $5000<t<6000$, an interval in which the statistics had converged.

Figure~\ref{fig:flow-sequences} compares the short-time evolution of the vorticity fields. Its four rows show, respectively, the full HF solution, the same solution projected onto the resolved modes, LF, and SM. The six columns follow each realization at times $t=0,2,4,6,8,10$. Colour represents the sign and magnitude of vorticity, a measure of local fluid rotation, using a common scale across all panels. Comparing the first two rows shows how projection removes fine-scale filaments while retaining the larger structures. The third and fourth rows show how those resolved structures evolve under the two reduced models. LF shares the displayed coarse HF initial state, whereas SM starts from an independent HF snapshot. The comparison therefore concerns the spatial structure and temporal evolution of the fields, without requiring the SM realization to track this particular HF trajectory.

Figure~\ref{fig:statistics} examines both typical fluctuations and departures from them. The top-left panel shows the mean and standard deviation of the energy in each Fourier shell: smaller $k$ corresponds to larger spatial structures, and the standard deviation measures variability about the mean. The top-middle panel compares the probability density functions (PDFs) of total energy $E$ and enstrophy $Z$, the latter measuring the spatially averaged squared vorticity. The top-right panel shows PDFs of $E_z/E$, the fraction of energy in the $k_x=0$ zonal modes, and of spatial vorticity kurtosis $K_\omega=\langle\omega^4\rangle_{x,y}/\langle\omega^2\rangle_{x,y}^2$, which measures the contribution of large vorticity values relative to the overall fluctuation amplitude. All PDFs use the HF mean and standard deviation for normalization. The middle-left panel shows second-order structure functions of velocity and vorticity. For either field $f$, an increment $\delta_r f=f(\boldsymbol{x}+r\boldsymbol{e})-f(\boldsymbol{x})$ is the difference between two points separated by $r$; $S_2(r)=\langle|\delta_r f|^2\rangle$ measures its mean squared size. The middle panel shows increment flatness $S_4(r)/S_2(r)^2$, where $S_4(r)=\langle|\delta_r f|^4\rangle$. Larger flatness means that large increments contribute more strongly relative to their mean squared size; Gaussian increments have flatness three. The middle-right panel shows the joint energy--enstrophy PDF, testing whether the models reproduce their simultaneous fluctuations. The bottom-left panel relates total energy to the fraction organized into zonal flow; the bottom-middle panel relates the neighbouring shell energies $E_3$ and $E_4$; and the bottom-right panel relates enstrophy to the concentration of large vorticity values through $K_\omega$. In these four joint PDFs, contour location describes typical pairs of values, contour extent describes their spread, and contour orientation and shape reveal statistical dependence that separate PDFs cannot show. The contours enclose 50\% and 90\% of the displayed probability.

Figure~\ref{fig:acf-comparison} tests how long fluctuations in these observables persist. Each curve is the covariance of an observable at times separated by $\tau$, divided by its variance: values near one indicate strong persistence, values near zero indicate little linear correlation, and negative values indicate a tendency for fluctuations of opposite sign. From left to right, the top row shows the autocorrelations of total energy, zonal energy fraction, enstrophy, and vorticity kurtosis. These panels respectively measure the persistence of overall energy fluctuations, changes in the partition between zonal and nonzonal flow, fluctuations in squared vorticity, and changes in the relative contribution of large vorticity values. The middle row resolves the four lowest shell energies, $E_1$ through $E_4$, while the first two panels of the bottom row show $E_8$ and $E_{16}$ to extend the comparison to smaller spatial scales. The bottom-row third and fourth panels show the autocorrelations of vorticity-increment flatness at fixed separation $\pi/16$ in the $x$ and $y$ directions, respectively. The time range shown in each panel is chosen to display the decay of its correlations. The shaded confidence intervals indicate the uncertainty of the ensemble estimates.

Figure~\ref{fig:cross-correlations} extends the energy comparison to correlations between different shells. Each column shows HF in the top row, SM in the middle row, and LF in the bottom row, at lags $\tau=0,1,10,100,500,1000$. Entry $(i,j)$ measures how fluctuations in the energy of shell $j$ are correlated with fluctuations in shell $i$ after a time interval $\tau$. The first column describes simultaneous fluctuations: its diagonal is unity, and its off-diagonal entries show which shells fluctuate together or oppositely. In the remaining columns, diagonal entries follow the shell autocorrelations, while off-diagonal entries describe the persistence and temporal ordering of inter-shell statistical dependence. Red entries indicate positive correlation, blue entries negative correlation, and near-white entries weak correlation on the common scale. Comparing the three rows tests whether the reduced models reproduce how long fluctuations in different shells remain related.

Figure~\ref{fig:jet-dynamics} examines zonal-flow organization and dynamical observables beyond shell energies. The three top panels show the zonally averaged velocity as a function of latitude and time for HF, LF, and SM, respectively. Here $U(y,t)=\langle u(x,y,t)\rangle_x$ is the east--west velocity averaged along the zonal direction. The coloured bands show the direction and strength of the jets, their positions in $y$, and how those positions and strengths change with time. The bottom-left panel compares the autocorrelation of the root-mean-square zonal shear, measuring how long changes in the variation of zonal velocity across $y$ persist. The bottom-middle panel shows the autocorrelation of the real part of the vorticity coefficient at wavevector $(1,1)$, which can change sign as the spatial phase of that Fourier pattern changes. Its correlation therefore contains information about the motion of the pattern that its squared amplitude, and hence its energy, cannot retain. In the QG equations, advection exchanges energy between eddies and zonal flow at a rate $P(t)=\langle\langle uv\rangle_x\,\partial_y U\rangle_y$. Here $u$ and $v$ are the zonal and meridional velocity components, $U=\langle u\rangle_x$, and the brackets denote spatial averages. Positive $P$ means that eddies supply energy to the zonal flow; negative $P$ means that the zonal flow supplies energy to the eddies. We evaluate this expression using the resolved velocity fields from HF, LF, and SM. For SM, $P$ is an observable calculated from the generated fields, rather than a term taken from its stochastic equations. The bottom-right panel compares the autocorrelation of $P$, testing whether the models reproduce the persistence of fluctuations in this QG energy-transfer quantity. Together, these panels compare jet variability, the evolution of a Fourier amplitude, and the temporal fluctuations of $P$.

\clearpage

%% file: appendix.tex
\clearpage
\appendix
\section{Numerical methods and calibration details}
\label{app:implementation}

\subsection{QG reference simulations}
\label{app:qg}

We refer to the full QG reference simulations as high-fidelity (HF) simulations. All quantities use the nondimensional units of Eq.~\eqref{eq:qg}. Spatial derivatives are evaluated in Fourier space and nonlinear products on the periodic physical grid. Each step applies fourth-order Runge--Kutta to the deterministic drift, followed by a Gaussian forcing increment. The zero Fourier mode is removed.

\begin{center}
\begin{tabular}{ll}
\hline
Parameter & Value \\
\hline
Domain; grid & $[0,2\pi)^2$; $512\times512$ \\
$\beta$; $\mu$; $\nu$ & $10$; $10^{-2}$; $1.03\times10^{-5}$ \\
Forcing energy injection rate $\varepsilon_f$ & $10^{-3}$ \\
Forcing centre $k_f$; bandwidth $\Delta k_f$ & $14$; $1.5$ \\
Time step; sampling interval & $0.004$; $0.2$ \\
Initial vorticity amplitude & $10^{-3}$ \\
Discarded transient; sampling duration & $3000$; $6000$ \\
Reference trajectories for normalization & $768$ \\
Integration; recorded-state precision & Double; single \\
Resolved cutoff; real-state dimension & $K=16$; $796$ \\
\hline
\end{tabular}
\end{center}

The forcing in Eq.~\eqref{eq:forcing-definition} uses $\mathcal K_f=\{\boldsymbol k\in\mathbb Z^2\setminus\{\boldsymbol0\}:|k_x|,|k_y|\leq170\}$. The filter $\exp[-36(|\boldsymbol{k}|/256)^{36}]$ is applied after the deterministic update and after forcing. Because the vorticity field is real, the coefficients at opposite wavevectors are complex conjugates, $\widehat\zeta_{-\boldsymbol k}=\widehat\zeta_{\boldsymbol k}^{\,*}$. We therefore retain one wavevector from each conjugate pair, defining the resolved set as
\begin{equation}
 \mathcal K_+=\{\boldsymbol k\in\mathbb Z^2:0<|\boldsymbol k|\leq16,
 \ k_x>0\ \text{or}\ (k_x=0,k_y>0)\}.
\end{equation}
The 398 retained complex coefficients give 796 real coordinates, ordered as all real parts followed by all imaginary parts. Using the unnormalized discrete Fourier transform on the $512^2$ grid,
\begin{equation}
 e_{\boldsymbol k}=\frac{|\widehat\zeta_{\boldsymbol k}|^2}{512^4|\boldsymbol k|^2},
 \qquad E_j=\sum_{\substack{\boldsymbol k\in\mathcal K_+\\j-1<|\boldsymbol k|\leq j}}e_{\boldsymbol k},
 \qquad E=\sum_{j=1}^{16}E_j.
\end{equation}
Each $e_{\boldsymbol k}$ includes the energy of both conjugate modes. The shell energy $E_j$ is the kinetic energy contained in the wavenumber band $j-1<|\boldsymbol k|\leq j$, and $E$ is the total resolved kinetic energy.

\subsection{Conventional low-fidelity closure}
\label{app:lf}

We integrate the projected QG equations~\eqref{eq:projected-qg}, replacing the unresolved tendency $\mathcal R_K$ by eddy-viscosity damping $-\mathcal D_K\overline\zeta$ and stochastic backscatter $\xi_{\mathrm{bs}}$. These terms represent the loss of resolved energy to unresolved motion and the return of energy from unresolved scales, respectively. We call the projected QG equations with these closure terms the low-fidelity (LF) model. The resolved wavevector set, physical parameters, and physical-forcing spectrum are those specified above; the physical forcing is projected without renormalizing its amplitude. Each time step uses explicit midpoint for the deterministic drift, followed by independent Gaussian increments for physical forcing and backscatter.

\begin{center}
\begin{tabular}{ll}
\hline
Parameter & Value \\
\hline
Nonlinear evaluation grid & $128\times128$ \\
Eddy-viscosity exponent $p$ & $4$ \\
Eddy-viscosity amplitude $\nu_{\mathrm e}$ & $3.22\times10^{-3}$ \\
Backscatter energy injection rate $\varepsilon_{\mathrm{bs}}$ & $1.75\times10^{-3}$ \\
Backscatter centre $k_{\mathrm{bs}}$; bandwidth $\Delta k_{\mathrm{bs}}$ & $10$; $2$ \\
Retained physical-forcing energy injection & $9.17\times10^{-4}$ \\
Time step; comparison time step & $0.02$; $0.01$ \\
\hline
\end{tabular}
\end{center}

The damping operator is given by Eq.~\eqref{eq:eddy-viscosity}. The backscatter is defined by Eqs.~\eqref{eq:backscatter-definition}--\eqref{eq:backscatter-amplitude}, with the parameters listed in the table. The damping amplitude and backscatter energy injection rate are fitted to the energy and enstrophy budgets of the projected reference trajectories.

\subsection{Stationary score estimation}
\label{app:score}

The training data are resolved Fourier states extracted from the high-fidelity (HF) reference simulations. Each state has 796 real coordinates, ordered as in Appendix~\ref{app:qg}. We standardize coordinate $x_i$ using its reference-data mean $\mu_i$ and standard deviation $\sigma_i$:
\begin{equation}
 z_i=\frac{x_i-\mu_i}{\sigma_i}.
\end{equation}
The learned density combines a Gaussian reference $q_G$ with a bounded neural correction $h_\theta$,
\begin{equation}
 p_\theta(\boldsymbol z)\propto q_G(\boldsymbol z)e^{h_\theta(\boldsymbol z)},
 \qquad q_G=\mathcal N(\boldsymbol z_{\mathrm{ref}},C_G+\eta^2I).
\end{equation}
Here $\boldsymbol z_{\mathrm{ref}}$ is the estimated mean of the standardized data, $C_G$ is their estimated covariance matrix, and $\eta$ is the standard deviation of the Gaussian noise used in denoising score matching. The term $\eta^2I$ accounts for this added noise. A neural network with scalar output $f_\theta$ defines the correction and the score in standardized coordinates:
\begin{equation}
 h_\theta(\boldsymbol z)=200\tanh\!\left[
 \frac{f_\theta(\boldsymbol z)-f_\theta(\boldsymbol z_{\mathrm{ref}})}{200}\right],
 \qquad s_\theta(\boldsymbol z)=\nabla_{\boldsymbol z}\log q_G(\boldsymbol z)+\nabla_{\boldsymbol z}h_\theta(\boldsymbol z).
\end{equation}

\begin{center}
\begin{tabular}{ll}
\hline
Parameter & Value \\
\hline
Denoising standard deviation $\eta$ & $0.02$ \\
Network; spatial grid & Periodic U-Net; $40\times40$ \\
Encoder widths & $24,48,96,192$ \\
Residual blocks per encoder/decoder level & $2$ \\
Convolution kernel; normalization groups & $3\times3$; $6$ \\
Activation; noise embedding dimension & SiLU; $256$ \\
Training states; trajectories & $81{,}920$; $640$ \\
Validation states; trajectories & $128$; $128$ \\
Batch & 32 states; 64 perturbed inputs \\
Optimizer; learning rate & AdamW; $5\times10^{-7}$ \\
Weight decay; gradient-norm bound & $10^{-5}$; $10$ \\
Adam coefficients; denominator offset & $(0.9,0.999)$; $10^{-8}$ \\
Exponential moving-average decay & $0.99$ \\
Training updates & $36{,}000$ \\
Training and score precision & Single \\
Downsampling; upsampling & Stride-two convolution; nearest neighbour \\
\hline
\end{tabular}
\end{center}

The retained Fourier coefficients are converted into six input fields on a $40\times40$ grid. This grid represents all modes up to $K=16$ without aliasing and accommodates the U-Net's successive downsampling to $20\times20$, $10\times10$, and $5\times5$ grids. The first three inputs are the resolved vorticity $\zeta(x,y)$, the streamfunction obtained from $\Delta\psi=\zeta$, and the zonal-mean velocity $\overline u(y)=-(2\pi)^{-1}\int_0^{2\pi}\partial_y\psi(x,y)\,\mathrm dx$. To construct the other three inputs, we first divide each vorticity coefficient $\widehat\zeta_{\boldsymbol k}$ by its typical amplitude $\sigma_{\boldsymbol k}=(\sigma_{\boldsymbol k,\mathrm{Re}}+\sigma_{\boldsymbol k,\mathrm{Im}})/2$, computed from the reference data, and then reconstruct the same three fields. This rescaling makes modes with smaller typical amplitudes more visible to the network.

Training uses denoising score matching (DSM). For each sampled state $\boldsymbol z$, we draw $\boldsymbol\epsilon\sim\mathcal N(0,I)$ and evaluate the score at both $\boldsymbol z+\eta\boldsymbol\epsilon$ and $\boldsymbol z-\eta\boldsymbol\epsilon$. These opposite (antithetic) perturbations give the loss
\begin{equation}
 \mathcal L_{\mathrm{DSM}}=
 \frac{1}{2(796)\eta^2}
 \mathbb E_{\boldsymbol z,\boldsymbol\epsilon}
 \sum_{\delta\in\{-1,1\}}
 \|\eta s_\theta(\boldsymbol z+\delta\eta\boldsymbol\epsilon)
       +\delta\boldsymbol\epsilon\|^2.
\end{equation}
The expectation is over training states and Gaussian perturbations. The factor 796 averages over coordinates. For each training batch, we sample 32 reference states and construct two noisy inputs from each state, $\boldsymbol z+\eta\boldsymbol\epsilon$ and $\boldsymbol z-\eta\boldsymbol\epsilon$. The batch loss is evaluated on these 64 noisy inputs. We maintain an exponential moving average of the network parameters using $\overline\theta\leftarrow0.99\overline\theta+0.01\theta$. Validation uses one fixed state from each of 128 trajectories and fixed antithetic perturbations.

\subsection{Diffusion and circulation estimation}
\label{app:mobility}

Calibration uses two sources of data: projected high-fidelity trajectories supply the temporal statistics to be matched, and samples from the learned density $p_\theta$ supply the stationary averages needed to evaluate the model predictions. We divide the reference trajectories into 48 for fitting, 24 for choosing regularization strengths, and 24 for assessing the selected coefficients. We first estimate the preliminary mobility $\widehat\Phi$, then fit the diffusion $D$, and finally fit the circulation $R$ with $D$ held fixed. The numerical settings are collected below.

\begin{center}
\begin{tabular}{ll}
\hline
Parameter & Value \\
\hline
Time origins per trajectory; time interval & $2048$; $[2000,5999.2]$ \\
Preliminary-mobility forward lags & $0,0.2,0.4,0.6$ \\
Diffusion penalty coefficient $\alpha$ & $10^{-7}$ \\
Energy-forcing interval $[q_f^-,q_f^+]$ & $[1.08,1.09]\times10^{-8}$ \\
Projector regularizer $\epsilon$ & $8.83\times10^{-15}$ \\
Symmetric-difference half-widths $h$ & $0.2,0.4,0.8$ \\
Relative circulation penalty strength $\lambda_R$ & $10^{-5}$ \\
Reduced-model time step & $0.008$ \\
\hline
\end{tabular}
\end{center}

The preliminary mobility estimate uses a quadratic fit at the four forward lags and an SVD cutoff of $10^{-3}$ relative to the largest singular value; uncertainties use 32 whole-trajectory bootstrap resamples.

Let $\langle\cdot\rangle_f$ denote an average over the reference fitting data. The diffusion library contains 398 standardized modal energies and seven hinge functions for each of the 16 zonal modes:
\begin{align}
 u_{\boldsymbol k}&=\frac{e_{\boldsymbol k}-\langle e_{\boldsymbol k}\rangle_f}
 {\operatorname{std}_f(e_{\boldsymbol k})},
 &h_{j\ell}(u)&=\max(u-\kappa_{j\ell},0),\\
 \kappa_{j\ell}&=\operatorname{Quantile}_{\ell/8,f}(u_{(0,j)}),
 &j&=1,\ldots,16,\quad\ell=1,\ldots,7.
\end{align}
Each hinge is orthogonalized against $1$, $u$, and the preceding hinges for the same mode, then normalized to unit mean square under $\langle\cdot\rangle_f$. Denoting the resulting functions by $\phi_{j\ell}$, the 510-component library is
\begin{equation}
 \Psi_0=\left(\{u_{\boldsymbol k}\}_{\boldsymbol k\in\mathcal K_+},
 \{\phi_{j\ell}(u_{(0,j)})\}_{j,\ell}\right)^{\mathsf T},
 \qquad \Psi=\Psi_0-\mathbb E_{p_\theta}\Psi_0.
\end{equation}
The shell- and total-energy auto- and symmetrized cross-correlations are fitted at the 63 distinct lags
\begin{equation}
 \mathcal T=\left\{0.2\operatorname{round}(10^{4j/71}):j=0,\ldots,71\right\},
 \qquad 0.2\leq t\leq2000.
\end{equation}

We draw samples directly from the learned density $p_\theta$ using Hamiltonian Monte Carlo (HMC), without integrating the reduced dynamical model. Sampling is performed in the coordinates
\begin{equation}
 \boldsymbol y=(C_G+\eta^2I)^{-1/2}
 (\boldsymbol z-\boldsymbol z_{\mathrm{ref}}),
\end{equation}
which give the Gaussian reference zero mean and identity covariance. HMC uses gradients of the log-density to propose moves, followed by a Metropolis acceptance step. To improve exploration, we run eight chains targeting
\begin{equation}
 p_b(\boldsymbol z)\propto q_G(\boldsymbol z)\exp[b h_\theta(\boldsymbol z)],
 \qquad b=0,1/7,\ldots,1.
\end{equation}
These densities connect the Gaussian reference at $b=0$ to the learned density at $b=1$. Exchanges of states between neighbouring chains are accepted with a Metropolis rule that preserves their target densities. Only states from the $b=1$ chain are used to estimate stationary averages.

We run 24 such groups of chains, half initialized from reference states and half from Gaussian samples. For each group, we discard 2048 sampling updates and retain 16,384 states from the $b=1$ chain. The groups are divided into 12 for fitting, six for choosing regularization, and six for assessment. Within each subset, all retained states are used for $B$ and $q_E(C)$; every sixteenth state is used for $Q(D)$.

Symmetry leaves 215 diagonal coefficients in each of $C$ and $\Lambda$, and 199 circulation coefficients. Diffusion is fitted with L-BFGS-B in logarithmic coordinates to enforce positivity.

The circulation fit uses 49 observables: the 17 shell and total energies, and the real and imaginary Fourier components at
\begin{equation}
 \mathcal P=\{(1,\pm3),(3,\pm2),(2,\pm3),(1,\pm4),
 (1,\pm5),(2,\pm2),(3,\pm3),(3,\pm1)\}.
\end{equation}
These modes are selected from the reference trajectories for their measurable phase dynamics. We rank nonzonal reflection groups using the strength of the antisymmetric lagged correlation between real and imaginary components, its agreement between fitting and development data, and the persistence of the Fourier-component correlations. Only antisymmetric signals exceeding three estimated standard errors contribute to the strength measure. The eight highest-ranked groups give the 16 modes above, with both members of each reflection pair retained. Their real and imaginary components constrain phase motion that energy observables cannot resolve.
For each observable $F_b$, the temporal target and weight are
\begin{align}
 Y_h(t)&=\frac{F(t+h)-F(t-h)}{2h},
 &\widehat Y&=Y_{0.2}+\tfrac12Y_{0.4}-\tfrac12Y_{0.8},\\
 V_b&=\langle\widehat Y_b^2\rangle_f,
 &w_b&=\begin{cases}(17V_b)^{-1},&b\leq17,\\(32V_b)^{-1},&b>17.\end{cases}
\end{align}
These weights make the fit invariant to a constant rescaling of any observable. The circulation penalty in Eq.~\eqref{eq:circulation-objective} is centred at the preliminary estimates $r_a^\Phi$. Its coefficient $\gamma$ is computed using the preliminary uncertainty scales $u_a^R$:
\begin{equation}
 \gamma=\lambda_R\frac{\operatorname{tr}(U H U)}{199},
 \qquad U=\operatorname{diag}(u_a^R).
\end{equation}
\begin{equation}
 H_{ac}=\left\langle\sum_b w_b A_{ba}A_{bc}\right\rangle_f.
\end{equation}
Thus $\lambda_R$ sets the relative penalty strength, while $\gamma$ is the coefficient multiplying the penalty in the main-text notation. Here $A_{ba}$ is defined in the main-text circulation fit.

Reduced-model trajectories use the symmetric splitting
$\mathcal T_{\Delta t/2}\mathcal B_{\Delta t}\mathcal T_{\Delta t/2}$.
The step $\mathcal T$ advances $D-C$ and $R$ by stochastic Heun with the Stratonovich correction, rescaling the predictor and corrected state to preserve energy. The step $\mathcal B$ advances the constant-diffusion contribution using exact Gaussian Ornstein--Uhlenbeck half-steps around a Heun step for the neural-score drift. States and mobilities use double precision, score evaluations use single precision, and states are recorded every $0.2$ time units.

%% file: main.bbl
\begin{thebibliography}{99}

\bibitem{Berner2017}
J. Berner et al.
\newblock Stochastic parameterization: Toward a new view of weather and climate models.
\newblock \emph{Bulletin of the American Meteorological Society}, \textbf{98}, 565--588 (2017).
\newblock \href{https://doi.org/10.1175/BAMS-D-15-00268.1}{doi:10.1175/BAMS-D-15-00268.1}.

\bibitem{WoutersLucarini2013}
J. Wouters and V. Lucarini.
\newblock Multi-level dynamical systems: Connecting the Ruelle response theory and the Mori--Zwanzig approach.
\newblock \emph{Journal of Statistical Physics}, \textbf{151}, 850--860 (2013).
\newblock \href{https://doi.org/10.1007/s10955-013-0726-8}{doi:10.1007/s10955-013-0726-8}.

\bibitem{FrederiksenDavies1997}
J.~S. Frederiksen and A.~G. Davies.
\newblock Eddy viscosity and stochastic backscatter parameterizations on the sphere for atmospheric circulation models.
\newblock \emph{Journal of the Atmospheric Sciences}, \textbf{54}, 2475--2492 (1997).
\newblock \href{https://doi.org/10.1175/1520-0469(1997)054<2475:EVASBP>2.0.CO;2}{doi:10.1175/1520-0469(1997)054\textless2475:EVASBP\textgreater2.0.CO;2}.

\bibitem{Frezat2022}
H. Frezat, J. Le Sommer, R. Fablet, G. Balarac and R. Lguensat.
\newblock A posteriori learning for quasi-geostrophic turbulence parametrization.
\newblock \emph{Journal of Advances in Modeling Earth Systems}, \textbf{14}, e2022MS003124 (2022).
\newblock \href{https://doi.org/10.1029/2022MS003124}{doi:10.1029/2022MS003124}.

\bibitem{BotvinickGreenhouse2023}
J. Botvinick-Greenhouse, R. Martin and Y. Yang.
\newblock Learning dynamics on invariant measures using PDE-constrained optimization.
\newblock \emph{Chaos}, \textbf{33}, 063152 (2023).
\newblock \href{https://doi.org/10.1063/5.0149673}{doi:10.1063/5.0149673}.

\bibitem{Giorgini2026Score}
L.~T. Giorgini.
\newblock Score-based modeling of effective Langevin dynamics.
\newblock \emph{Physical Review E}, \textbf{114}, L012102 (2026).
\newblock \href{https://doi.org/10.1103/6qpv-lqmt}{doi:10.1103/6qpv-lqmt}.

\bibitem{Giorgini2026Conditional}
L.~T. Giorgini.
\newblock Conditional Score-Based Modeling of Effective Langevin Dynamics.
\newblock arXiv:2604.23952 (2026).
\newblock \url{https://arxiv.org/abs/2604.23952}.

\bibitem{FalascaZanna2026}
F. Falasca and L. Zanna.
\newblock Physics constraints and response validation in discrete-time reduced-order modeling: from idealized turbulent systems to climate dynamics.
\newblock arXiv:2602.13847v5 (2026).
\newblock \url{https://arxiv.org/abs/2602.13847v5}.

\bibitem{Song2021}
Y. Song, J. Sohl-Dickstein, D.~P. Kingma, A. Kumar, S. Ermon and B. Poole.
\newblock Score-based generative modeling through stochastic differential equations.
\newblock \emph{International Conference on Learning Representations} (2021).
\newblock \url{https://arxiv.org/abs/2011.13456}.

\bibitem{Marconi2008}
U. Marini Bettolo Marconi, A. Puglisi, L. Rondoni and A. Vulpiani.
\newblock Fluctuation--dissipation: Response theory in statistical physics.
\newblock \emph{Physics Reports}, \textbf{461}, 111--195 (2008).
\newblock \href{https://doi.org/10.1016/j.physrep.2008.02.002}{doi:10.1016/j.physrep.2008.02.002}.

\bibitem{Giorgini2024Response}
L.~T. Giorgini, K. Deck, T. Bischoff and A. Souza.
\newblock Response theory via generative score modeling.
\newblock \emph{Physical Review Letters}, \textbf{133}, 267302 (2024).
\newblock \href{https://doi.org/10.1103/PhysRevLett.133.267302}{doi:10.1103/PhysRevLett.133.267302}.

\bibitem{Giorgini2025Response}
L.~T. Giorgini, F. Falasca and A.~N. Souza.
\newblock Predicting forced responses of probability distributions via the fluctuation--dissipation theorem and generative modeling.
\newblock \emph{Proceedings of the National Academy of Sciences}, \textbf{122}, e2509578122 (2025).
\newblock \href{https://doi.org/10.1073/pnas.2509578122}{doi:10.1073/pnas.2509578122}.

\bibitem{MarstonTobias2023}
J.~B. Marston and S.~M. Tobias.
\newblock Recent developments in theories of inhomogeneous and anisotropic turbulence.
\newblock \emph{Annual Review of Fluid Mechanics}, \textbf{55}, 351--375 (2023).
\newblock \href{https://doi.org/10.1146/annurev-fluid-120720-031006}{doi:10.1146/annurev-fluid-120720-031006}.

\bibitem{FrederiksenKitsiosOKane2024}
J.~S. Frederiksen, V. Kitsios and T.~J. O'Kane.
\newblock Statistical Dynamics and Subgrid Modelling of Turbulence: From Isotropic to Inhomogeneous.
\newblock \emph{Atmosphere}, \textbf{15}, 921 (2024).
\newblock \href{https://doi.org/10.3390/atmos15080921}{doi:10.3390/atmos15080921}.

\bibitem{Klus2020}
S. Klus, F. N\"uske, S. Peitz, J.-H. Niemann, C. Clementi and C. Sch\"utte.
\newblock Data-driven approximation of the Koopman generator: Model reduction, system identification, and control.
\newblock \emph{Physica D: Nonlinear Phenomena}, \textbf{406}, 132416 (2020).
\newblock \href{https://doi.org/10.1016/j.physd.2020.132416}{doi:10.1016/j.physd.2020.132416}.

\bibitem{GalletFerrari2021}
B. Gallet and R. Ferrari.
\newblock A quantitative scaling theory for meridional heat transport in planetary atmospheres and oceans.
\newblock \emph{AGU Advances}, \textbf{2}, e2020AV000362 (2021).
\newblock \href{https://doi.org/10.1029/2020AV000362}{doi:10.1029/2020AV000362}.

\bibitem{QiMajda2016}
D. Qi and A.~J. Majda.
\newblock Low-dimensional reduced-order models for statistical response and uncertainty quantification: Two-layer baroclinic turbulence.
\newblock \emph{Journal of the Atmospheric Sciences}, \textbf{73}, 4609--4639 (2016).
\newblock \href{https://doi.org/10.1175/JAS-D-16-0192.1}{doi:10.1175/JAS-D-16-0192.1}.

\end{thebibliography}
